\documentclass[twocolumn,showpacs,aps,prd,floatfix]{revtex4}

\usepackage{epsfig}
\usepackage{graphics}
\usepackage{graphicx}
\usepackage{epic}
\usepackage{floatflt}
\usepackage{dcolumn}
\usepackage{amsmath}
\usepackage{colordvi}

\input babarsym
\def\babar{\mbox{\slshape B\kern-0.1em{\smaller A}\kern-0.1em
    B\kern-0.1em{\smaller A\kern-0.2em R}}\xspace}

\def\CL {\ensuremath{ \rm C.L. }\xspace}
\def\twoDLL  {\ensuremath{2\Delta\ln{\cal L}}\xspace}
\def\Bm {\ensuremath{B^-}\xspace}
\def\Bp {\ensuremath{B^+}\xspace}

\def\Dp {\ensuremath{D^+}\xspace}

\def\Dmp {\ensuremath{D^\mp}\xspace}
\def\rhopm {\ensuremath{\rho(770)^\pm}\xspace}
\def \uvec {\ensuremath {{\mathbf u}}\xspace}

\def \pdf {{P.D.F.}\xspace}
\def \pdfs {{P.D.F.s}\xspace}

\def\beq{\begin{equation}}
\def\eeq{\end{equation}}
\def\bea{\begin{eqnarray}}
\def\eea{\end{eqnarray}}
\def\bq{\begin{quote}}
\def\eq{\end{quote}}
\def\ben{\begin{enumerate}}
\def\een{\end{enumerate}}
\def\nn{\nonumber}

\newcommand{\phm}{\ensuremath{\phantom{-}}}

\newcommand{\phz}{\ensuremath{\phantom{0}}}
\newcommand{\phzz}{\ensuremath{\phantom{00}}}

\def\kspipi{\ensuremath{\KS \pip \pim}\xspace}

\def\kskk{\ensuremath{\KS \Kp \Km}\xspace}

\def\D {\ensuremath{D}\xspace}
\def\K {\ensuremath{K}\xspace}

\def\KKstarpm   {\ensuremath{K^{(\ast)\pm}}\xspace}

\def\DDstarz {\ensuremath {\D^{(\ast)0}}\xspace}
\def\DDstarzb {\ensuremath {\Db^{(\ast)0}}\xspace}
\def\DDstar   {\ensuremath{D^{(\ast)}}\xspace}

\def\DDstarmp   {\ensuremath{D^{(\ast)\mp}}\xspace}
\def\splus{\ensuremath{m^2_{\rm +}}\xspace}
\def\sminus{\ensuremath{m^2_{\rm -}}\xspace}

\def\splusminus{\ensuremath{m^2_{\rm \pm}}\xspace}
\def\sminusplus{\ensuremath{m^2_{\rm \mp}}\xspace}
\def\hp{\ensuremath{h^{+}}\xspace}
\def\hm{\ensuremath{h^{-}}\xspace}

\def\h{\ensuremath{h}\xspace}
\def \rb {\ensuremath {r_\B}\xspace}
\def \rbpm {\ensuremath {r_{\Bpm}}\xspace}

\def \rbst {\ensuremath {r^\ast_\B}\xspace}
\def \rbrbst {\ensuremath {r^{(\ast)}_\B}\xspace}
\def \rbrbstpm {\ensuremath {r^{(\ast)}_{\Bpm}}\xspace}

\def \rs {\ensuremath {r_s}\xspace}
\def \rspm {\ensuremath {r_{s\pm}}\xspace}

\def \krs {\ensuremath {\kappa \rs}\xspace}
\def \deltab {\ensuremath {\delta_\B}\xspace}
\def \deltabst {\ensuremath {\delta^\ast_\B}\xspace}
\def \deltabdeltabst {\ensuremath {\delta^{(\ast)}_\B}\xspace}
\def \deltas {\ensuremath {\delta_s}\xspace}

\def \xbpm {\ensuremath {{x}_\pm}\xspace}

\def \ybpm {\ensuremath {{y}_\pm}\xspace}

\def \zbp {\ensuremath {{\mathsf z}_+}\xspace}
\def \zbm {\ensuremath {{\mathsf z}_-}\xspace}

\def \zbpmtrue {\ensuremath {\overline{\mathsf z}_\pm}\xspace}
\def \zbptrue {\ensuremath {\overline{\mathsf z}_+}\xspace}
\def \zbmtrue {\ensuremath {\overline{\mathsf z}_-}\xspace}
\def \xbpmtrue {\ensuremath {\overline{x}_\pm}\xspace}
\def \xbptrue {\ensuremath {\overline{x}_+}\xspace}
\def \xbmtrue {\ensuremath {\overline{x}_-}\xspace}
\def \ybpmtrue {\ensuremath {\overline{y}_\pm}\xspace}

\def \zbstp {\ensuremath {{\mathsf z}_+^\ast}\xspace}
\def \zbstm {\ensuremath {{\mathsf z}_-^\ast}\xspace}

\def \zbstpmtrue {\ensuremath {\overline{\mathsf z}_\pm^\ast}\xspace}
\def \xbstpmtrue {\ensuremath {\overline{x}_\pm^\ast}\xspace}
\def \ybstpmtrue {\ensuremath {\overline{y}_\pm^\ast}\xspace}
\def \zbstptrue {\ensuremath {\overline{\mathsf z}_+^\ast}\xspace}
\def \zbstmtrue {\ensuremath {\overline{\mathsf z}_-^\ast}\xspace}

\def \xbxbstp {\ensuremath {{x}_+^{(\ast)}}\xspace}
\def \xbxbstm {\ensuremath {{x}_-^{(\ast)}}\xspace}
\def \xbxbstpm {\ensuremath {{x}_\pm^{(\ast)}}\xspace}

\def \ybybstpm {\ensuremath {{y}_\pm^{(\ast)}}\xspace}

\def \zbzbst {\ensuremath {{\mathsf z}^{(\ast)}}\xspace}
\def \zbzbstp {\ensuremath {{\mathsf z}_+^{(\ast)}}\xspace}
\def \zbzbstm {\ensuremath {{\mathsf z}_-^{(\ast)}}\xspace}
\def \zbzbstpm {\ensuremath {{\mathsf z}_\pm^{(\ast)}}\xspace}

\def \zbzbst {\ensuremath {{\mathsf z}^{(\ast)}}\xspace}

\def \xbxbstpmtrue {\ensuremath {\overline{x}_\pm^{(\ast)}}\xspace}
\def \ybybstpmtrue {\ensuremath {\overline{y}_\pm^{(\ast)}}\xspace}
\def \zbzbstpmtrue {\ensuremath {\overline{\mathsf z}_\pm^{(\ast)}}\xspace}

\def \xspm {\ensuremath {{x}_{s\pm}}\xspace}

\def \yspm {\ensuremath {{y}_{s\pm}}\xspace}

\def \zsp {\ensuremath {{\mathsf z}_{s+}}\xspace}
\def \zsm {\ensuremath {{\mathsf z}_{s-}}\xspace}
\def \zspm {\ensuremath {{\mathsf z}_{s\pm}}\xspace}

\def \zspmtrue {\ensuremath {\overline{\mathsf z}_{s\pm}}\xspace}
\def \xspmtrue {\ensuremath {\overline{x}_{s\pm}}\xspace}
\def \yspmtrue {\ensuremath {\overline{y}_{s\pm}}\xspace}
\def \zsptrue {\ensuremath {\overline{\mathsf z}_{s+}}\xspace}
\def \zsmtrue {\ensuremath {\overline{\mathsf z}_{s-}}\xspace}
\newcommand{\BABARPubYear}    {12}
\newcommand{\BABARPubNumber}  {025}

\newcommand{\SLACPubNumber} {15328}

\def\figurebox#1#2#3{%
    \def\arg{#3}%
    \ifx\arg\empty
    {\hfill\vbox{\hsize#2\hrule\hbox to #2{\vrule\hfill\vbox to #1{\hsize#2\vfill}\vrule}\hrule}\hfill}%
    \else
    {\hfill\epsfbox{#3}\hfill}%
    \fi}

\def \rd {\ensuremath {r_\D}\xspace}
\def \rKpipiz {\ensuremath {r_{\K\pi\piz}}\xspace}
\def \deltad {\ensuremath {\delta_\D}\xspace}
\def \kappaKpipiz {\ensuremath {\kappa_{\K\pi\piz}}\xspace}
\def \deltaKpipiz {\ensuremath {\delta_{\K\pi\piz}}\xspace}

\begin{document}

\begin{flushleft}
\babar-PUB-\BABARPubYear/\BABARPubNumber\\
SLAC-PUB-\SLACPubNumber   \\
\end{flushleft}

\title{
\boldmath
Observation of direct \CP violation in the measurement of the Cabibbo-Kobayashi-Maskawa angle \g with $\Bpm \to \DDstar \KKstarpm$ decays
}

%
\author{J.~P.~Lees}
\author{V.~Poireau}
\author{V.~Tisserand}
\affiliation{Laboratoire d'Annecy-le-Vieux de Physique des Particules (LAPP), Universit\'e de Savoie, CNRS/IN2P3,  F-74941 Annecy-Le-Vieux, France}
\author{E.~Grauges}
\affiliation{Universitat de Barcelona, Facultat de Fisica, Departament ECM, E-08028 Barcelona, Spain }
\author{A.~Palano$^{ab}$ }
\affiliation{INFN Sezione di Bari$^{a}$; Dipartimento di Fisica, Universit\`a di Bari$^{b}$, I-70126 Bari, Italy }
\author{G.~Eigen}
\author{B.~Stugu}
\affiliation{University of Bergen, Institute of Physics, N-5007 Bergen, Norway }
\author{D.~N.~Brown}
\author{L.~T.~Kerth}
\author{Yu.~G.~Kolomensky}
\author{G.~Lynch}
\affiliation{Lawrence Berkeley National Laboratory and University of California, Berkeley, California 94720, USA }
% ...
\collaboration{The \babar\ Collaboration}
\noaffiliation

\begin{abstract}\noindent
We report the determination of the Cabibbo-Kobayashi-Maskawa
\CP-violating angle \g through the combination of various measurements
involving $\Bpm \to \D \Kpm$, $\Bpm \to \Dstar \Kpm$, and $\Bpm \to \D
\Kstarpm$ decays performed by the \babar experiment at the \pep2
\epem\ collider at SLAC National Accelerator Laboratory.
Using up to 474 million \BB pairs, we obtain $\g = (69 ^{+17}_{-16})^\circ$
\mbox{modulo $180^\circ$}. The total uncertainty is dominated by
the statistical component, with the experimental and amplitude-model
systematic uncertainties amounting to $\pm 4^\circ$. 
The corresponding two-standard-deviation region is $41^\circ < \gamma
< 102^\circ$. 
This result is inconsistent with $\g = 0$ with a significance of $5.9$
standard deviations.
\end{abstract}


\maketitle

\section{ \bf Introduction and Overview}

In the Standard Model (SM), the mechanism of \CP violation in weak interactions 
arises from the joint effect of three mixing angles and the 
single irreducible phase in the three-family Cabibbo-Kobayashi-Maskawa (CKM) 
quark-mixing matrix~\cite{ref:ckm}. 
The unitarity of the CKM matrix $V$ implies a set of relations
among its elements, $V_{ij}$, with $i=u,c,t$ and $j=d,s,b$.
In particular, $V_{ud}V^*_{ub}+V_{cd}V^*_{cb}+V_{td}V^*_{tb} = 0$,
which can be depicted in the complex plane as a unitarity triangle
whose sides and angles are related to the magnitudes and phases
of the six elements of the first 
and third 
columns of the matrix,
$V_{id}$ and $V_{ib}$. 
The parameter \g, defined as $\arg [-V_{ud}V_{ub}^*/V_{cd}V_{cb}^*]$,
is one of the three angles of this
triangle. From measurements of the sides and
angles of the unitarity triangle
from many decay processes, 
it is possible to overconstrain our knowledge of the CKM mechanism, probing
dynamics beyond the 
SM~\cite{ref:globalCKMfits}. 
In this context, 
the angle \g is particularly relevant since it is the only \CP-violating parameter 
that can be cleanly determined using tree-level \B meson decays \cite{ref:Zupan}. 
In spite of a decade of successful operation and experimental efforts by
the $B$ factory experiments, \babar and Belle,
\g is poorly known due to its large statistical uncertainty.
Its precise determination is an important goal of present and future
flavor physics experiments.

\begin{figure}[htb!]
\includegraphics[width=0.23\textwidth]{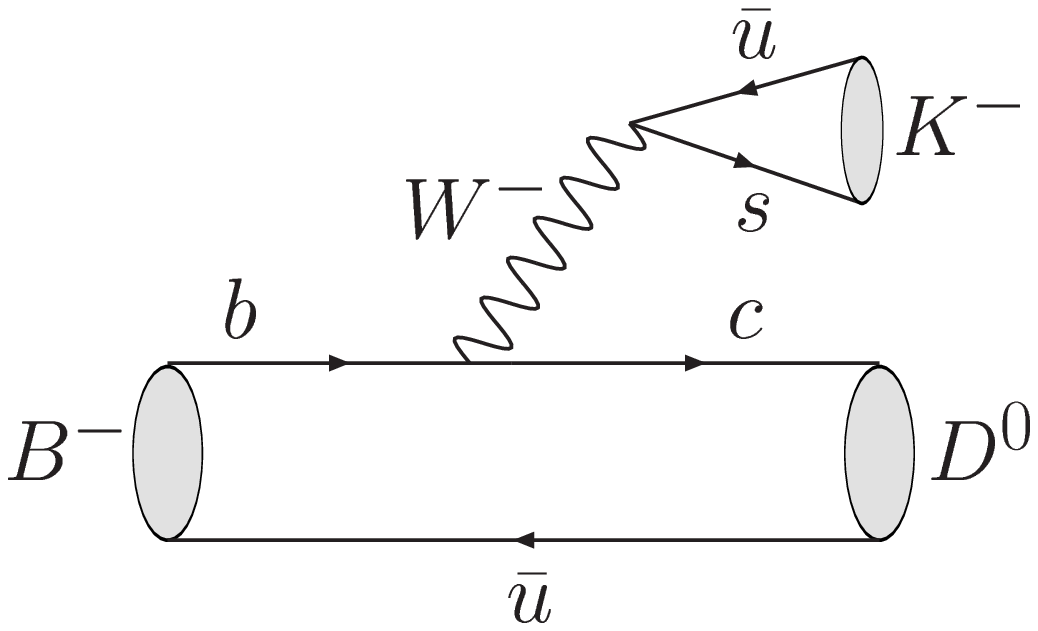}
\includegraphics[width=0.23\textwidth]{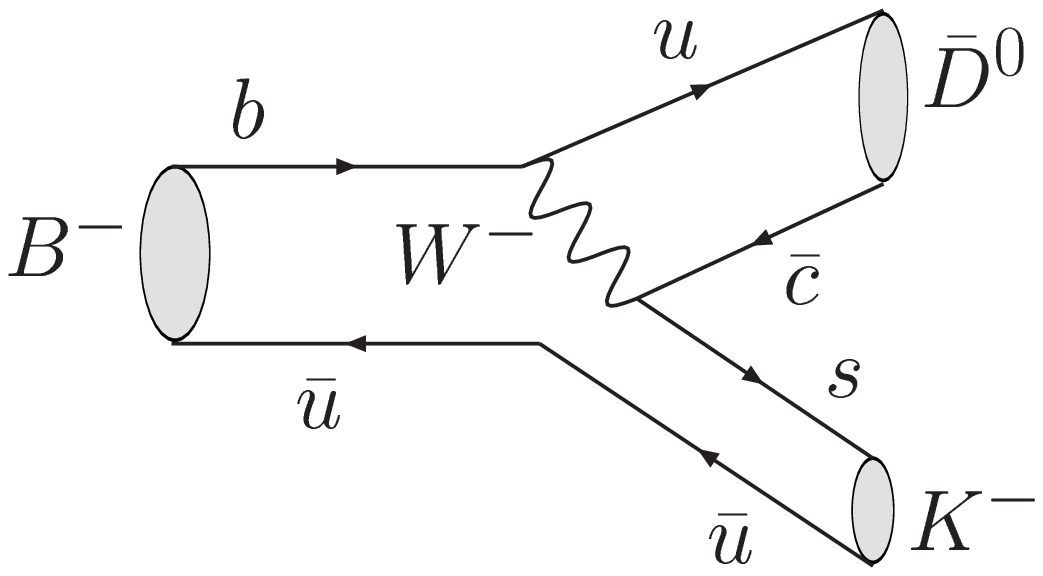}
    \caption{\label{fig:feynman_diagrams} 
Dominant Feynman diagrams for the decays $\Bm\to \Dz\Km$ (left) and $\Bm\to\Dzb\Km$ (right).
The left diagram proceeds via $b \to c \ubar  s$ transition, while the right diagram proceeds via 
       $b \to u \cbar s$ transition and is both CKM- and color-suppressed.}
  \end{figure}

Several methods have been pursued to extract \g~\cite{cite:gammaB2hh,cite:gammaB2hhh,cite:gammaB2hhh2,ref:dalitz_theo,ref:glw_theo,ref:ads_theo}. 
Those using charged \B meson decays into 
$\DDstar\Kpm$ and $\D\Kstarpm$ final states,
denoted generically as $\DDstar\KKstarpm$,
yield low theoretical uncertainties since the decays involved
do not receive contributions from penguin diagrams (see Fig.~\ref{fig:feynman_diagrams}).
This is a very important distinction from most other measurements of the angles.
Here, the symbol \DDstar indicates either a \Dz (\Dstarz) or a \Dzb (\Dstarzb) meson, and
\Kstarpm refers to $\Kstar(892)^\pm$ states. 
The methods to measure \g based on $\Bpm\to \DDstar\KKstarpm$ decays
rely on the interference between the CKM- and
color-favored $b \to c \ubar s$ and the 
suppressed $b\to u \cbar s$ amplitudes, 
which arises when the \Dz from a $\Bm \to \Dz \Km$ decay~\cite{ref:charge} (and similarly for the other related \B decays)
is reconstructed in a final state which can be produced also in the
decay of a $\Dzb$ originating from $\Bm \to \Dzb \Km$ 
(see Fig~\ref{fig:feynman_diagrams}).
The interference between the $b\to c\ubar s$ and
$b \to u \cbar s$ tree amplitudes results in observables that depend
on their relative weak phase \g, on 
the magnitude ratio $\rb \equiv | {\cal A}(b\to u\cbar s) / {\cal A}(b\to c\ubar s) |$
and on the relative strong phase $\deltab$ between the two amplitudes.
In the case of a nonzero weak phase \g and a nonzero
strong phase $\deltab$, the \Bm and \Bp decay rates are different,
a manifestation of direct \CP violation.
The hadronic parameters \rb and \deltab are not precisely known from 
theory, and may have different values for 
$\D\Kpm$, $D^*\Kpm$, and $\D\Kstarpm$ final states.
They can be measured directly from data by simultaneously
reconstructing several \D decay final states. 

The three main approaches employed by the $B$ factory experiments are:
\begin{itemize}
\item the Dalitz plot or Giri-Grossman-Soffer-Zupan (GGSZ) method,
based on three-body, self-conjugate final states, such as \kspipi~\cite{ref:dalitz_theo};

\item the Gronau-London-Wyler (GLW) method, 
based on decays to \CP-eigenstate final states, such as $\Kp \Km$ and $\KS\piz$~\cite{ref:glw_theo};

\item the Atwood-Dunietz-Soni (ADS) method,
based on \D decays to doubly-Cabibbo-suppressed final states,
such as $\Dz \to \Kp\pim$~\cite{ref:ads_theo}.
\end{itemize}

To date, the GGSZ method has provided the highest statistical power in measuring \g. The other two
methods provide additional information that 
can further constrain the hadronic parameters and thus allow for a more robust determination of \g. 
The primary issue with all these methods is the small product branching fraction of the decays involved, 
which range from $5\times10^{-6}$ to $5\times 10^{-9}$, and the small size of the interference,
proportional to $\rb \approx c_F | V_{cs} V_{ub}^*| / |V_{us} V_{cb}^*| \approx 0.1$, where $c_F \approx 0.2$ is a color suppression
factor~\cite{ref:browder1996,ref:gronau2003,ref:colsupBaBar}. 
Therefore a precise determination of \g requires a very large data sample and the combination of all available methods 
involving
different \D decay modes. 

Recently, Belle~\cite{cite:gammaBelle} and LHCb~\cite{cite:gammaLHCb} have presented the
preliminary
results of the combination of their measurements related to $\gamma$, yielding \g to be 
$\left(68^{+15}_{-14}\right)^{\circ}$ and $\left(71^{+17}_{-16}\right)^{\circ}$, respectively. Attempts to combine the results by 
\babar, Belle, CDF, and LHCb have been performed by the CKMfitter and UTfit groups~\cite{ref:globalCKMfits}. 
Their most recent results are $(66\pm12)^{\circ}$ and $(72\pm9)^{\circ}$, respectively

The \babar\ experiment~\cite{ref:detector} at the \pep2 asymmetric-energy \epem\ collider at SLAC
has analyzed charged \B decays into $\D\Kpm$, $\Dstar \Kpm$, and $\D\Kstarpm$ final states
using the GGSZ~\cite{ref:GGSZ2010,ref:GGSZ2008,ref:GGSZ2005}, 
GLW~\cite{ref:GLW_d0k,ref:GLW_dstar0k,ref:GLWADS_d0kstar} 
and 
ADS~\cite{ref:ADS_d0k_dstar0k_kpi,ref:ADS_d0k_kpipi0,ref:GLWADS_d0kstar} 
methods, providing a variety of measurements and constraints on \g.
The results are based on a dataset 
collected at a center-of-mass energy equal to the mass of the
\FourS resonance, and 
about 10\% of data collected 40 MeV below.
We present herein
the combination of published \babar\ measurements 
using detailed information on correlations between parameters that we have not 
previously published.
This combination represents the most complete study of the data sample 
collected by \babar\ and benefits from the possibility to access and
reanalyze the data sample (see Sec.~\ref{sec:methods} for detail). 

Other analyses related to $\gamma$~\cite{ref:GGSZ_pipipi0,ref:GGSZ_d0kstar0,ref:ADS_d0kstar0}
or $2\beta+\gamma$~\cite{ref:Dstpi_partialreco,ref:DpiDstpiDrho_fullreco} have not been 
included, because the errors on the experimental measurements are too large.

\section{ \bf Input Measurements\label{sec:methods}}


In the GGSZ approach, where \D mesons are reconstructed into the
\kspipi and \kskk final states~\cite{ref:GGSZ2010,ref:GGSZ2008,ref:GGSZ2005}, the signal rates 
for $\Bpm \to \DDstar \Kpm$ and $\Bpm \to \D \Kstarpm$ decays
are analyzed as a function of the position in the Dalitz plot of squared invariant masses 
$\sminus = m^2(\KS \hm)$, $\splus = m^2(\KS \hp)$, where \h 
is either a charged pion or kaon ($h = \pi, \K$).
We assume no \CP violation in the neutral \D and \K meson systems
and neglect small $\Dz-\Dzb$ mixing effects~\cite{ref:hfag2012,ref:mixingeffects}, 
leading to
$\overline{{\cal A}}(\sminus,\splus) = {\cal A}(\splus,\sminus)$, 
where $\overline{{\cal A}}$ (${\cal A}$) is the \Dzb (\Dz) decay amplitude.
In this case, the signal decay rates can be written as~\cite{ref:note_on_GGSZrates}
\bea
\label{eq:GGSZrates}
&&\Gamma^{(*)}_{\pm}(\sminus,\splus) \propto\nn\\
&&\phantom{\Gamma^{(*)}_{\pm}(\sminus)}|{\cal A}_{\pm}|^2+{\rbrbstpm}^2 |{\cal A}_{\mp}|^2 + 2 \lambda\operatorname{Re}[\zbzbstpm {\cal A}^\dagger_{\pm} {\cal A}_{\mp}],~\nn \\
&&\Gamma^s_{\pm}(\sminus,\splus)     \propto \nn\\
&&\phantom{\Gamma^{(*)}_{\pm}(\sminus)}|{\cal A}_{\pm}|^2+\rspm^2 |{\cal A}_{\mp}|^2 + 2 \operatorname{Re}[\zspm {\cal A}^\dagger_{\pm} {\cal A}_{\mp}],\label{eq:ampgen}
\eea
with ${\cal A}_{\pm} \equiv {\cal A}(\splusminus,\sminusplus)$ and ${\cal A}^\dagger_\pm$ is the complex conjugate of ${\cal A}_\pm$.  
The symbol $\lambda$ for $\Bpm\to\Dstar\Kpm$ accounts for the different \CP parity of the \Dstar when it is reconstructed into 
$\D\piz$ ($\lambda=+1$) and $\D\gamma$ ($\lambda=-1$) final states, as a consequence of the opposite
\CP eigenvalue of the \piz and the photon~\cite{ref:bondar_gershon}.
Here, \rbrbstpm and $\rspm$ are the magnitude ratios
between the $b\to u \cbar s$ and $b \to c \ubar s$ amplitudes 
for $B^\pm\to D^{(*)}K^{\pm}$ and $B^\pm\to D K^{*\pm}$ decays, respectively, 
and \deltabdeltabst, \deltas are their relative strong phases. 
The analysis extracts the \CP-violating observables~\cite{ref:GGSZ2005}
\begin{eqnarray}
\zbzbstpm & \equiv & \xbxbstpm + i \ybybstpm, \nn \\
\zspm     & \equiv & \xspm + i \yspm\label{eq:cartesian},
\end{eqnarray}
defined as the suppressed-to-favored complex amplitude ratios 
$\zbzbstpm=\rbrbstpm e^{i(\deltabdeltabst \pm \g)}$ and 
$\zspm = \kappa \rspm e^{i(\deltas \pm \g)}$,
for $B^\pm\to D^{(*)}K^{\pm}$ and $B^\pm\to D K^{*\pm}$ decays,
respectively.
The hadronic parameter $\kappa$ is defined as
\bea
\kappa e^{i\delta_{s}} \equiv \frac{\int A_c(p) A_u(p) e^{i\delta(p)}{\rm d}p}{\sqrt{\int A^2_c(p){\rm d}p \int A^2_u(p){\rm d}p}},
\label{eq:kappa}
\eea
where $A_c(p)$ and $A_u(p)$ are the magnitudes of the $b \to c \ubar s$ and $b\to u \cbar s$ amplitudes as a function of 
the $\Bpm \to \D \KS \pipm$ phase space position $p$, and $\delta(p)$ is their relative strong phase. 
This coherence factor, with $0<\kappa<1$ in the most general case and $\kappa=1$ for two-body \B decays,
accounts for the interference between $\Bpm \to\D \Kstarpm$ and other $\Bpm \to\D \KS\pi^\pm$ decays,
as a consequence of the \Kstarpm natural width~\cite{ref:gronau2003}.
In our analysis, $\kappa$ has been fixed to $0.9$ and a 
systematic uncertainty has been assigned varying its value by $\pm0.1$,
as estimated using a Monte Carlo simulation based on Dalitz
plot model of $\Bpm \to\D \KS\pi^\pm$ decays~\cite{ref:GGSZ2008}.
Thus, the parameter $\delta_{s}$ is an effective strong-phase 
difference averaged over the phase space.

\begin{table}[!htb]
\caption{\label{tab:GGSZresults}
\CP-violating complex parameters 
$\zbzbstpm \equiv \xbxbstpm + i \ybybstpm$
and 
$\zspm \equiv \xspm + i \yspm$,
measured using the GGSZ technique~\cite{ref:GGSZ2010}.
The first uncertainty is statistical, the second is the experimental systematic uncertainty and the third is 
the systematic uncertainty associated with the \Dz decay amplitude models. The sample analyzed contains 468 million \BB pairs.}
\begin{center}
\begin{ruledtabular}
\begin{tabular}{lrr}
           & Real part (\%)\phzz\phzz\phz &  Imaginary part (\%)\phzz \\ [0.025in] \hline 
 $\zbm$    & $\phz\phm6.0\pm3.9\pm0.7\pm0.6\phzz$    & $\phz\phm6.2\pm\phz4.5\pm0.4\pm0.6\phz$ \\
 $\zbp$    & $-10.3\pm3.7\pm0.6\pm0.7\phzz$          & $\phz-2.1\pm\phz4.8\pm0.4\pm0.9\phz$ \\
 $\zbstm$  & $-10.4\pm5.1\pm1.9\pm0.2\phzz$          & $\phz-5.2\pm\phz6.3\pm0.9\pm0.7\phz$    \\
 $\zbstp$  & $\phm14.7\pm5.3\pm1.7\pm0.3\phzz$       & $\phz-3.2\pm\phz7.7\pm0.8\pm0.6\phz$ \\
 $\zsm$    & $\phz\phm7.5\pm9.6\pm2.9\pm0.7\phzz$    & $\phm12.7\pm\phz9.5\pm2.7\pm0.6\phz$    \\
 $\zsp$    & $-15.1\pm8.3\pm2.9\pm0.6\phzz$          & $\phz\phm4.5\pm10.6\pm3.6\pm0.8\phz$    \\
\end{tabular}
\end{ruledtabular}
\end{center}
\end{table}

Table~\ref{tab:GGSZresults} summarizes our experimental results for the \CP-violating parameters \zbzbstpm and \zspm. Complete
$12\times12$ covariance matrices for statistical, experimental systematic and amplitude model
uncertainties are reported in Ref.~\cite{ref:GGSZ2010}. 
The \zbzbstpm and \zspm observables 
are unbiased and have Gaussian behavior with small correlations,
even for low values of \rbrbst, $\kappa \rs$
and relatively low statistics samples. Furthermore, their uncertainties have minimal dependence 
on their central values
and are free of physical bounds~\cite{ref:GGSZ2005}.
These good statistical properties allow for easier combination of several measurements into a single result.
For example, the rather complex experimental GGSZ likelihood function 
can be parameterized by a 12-dimensional (correlated) Gaussian probability density function (\pdf),
defined in the space of the \zbzbstpm and \zspm measurements from Table~\ref{tab:GGSZresults}.
After this combination has been performed, the values 
of \g and of the hadronic parameters \rbrbst, \krs, \deltabdeltabst, and \deltas can be obtained.

The \D decay amplitudes ${\cal A}_{\pm}$
have been determined from Dalitz plot analyses of tagged \Dz mesons from $D^{*+} \to \Dz \pip$ decays produced 
in $e^+ e^- \to c\bar c$ events~\cite{ref:dmixing-kshh,ref:GGSZ2008}, assuming an empirical model to describe the 
variation of the amplitude phase as a function of the Dalitz plot variables.
A model independent, binned approach also exists~\cite{ref:dalitz_theo,ref:bondar_Poluektov}, which 
optimally extracts information on \g for higher statistics samples than the ones available.
This type of analysis has been performed as a proof of principle by the Belle collaboration~\cite{ref:belle_GGSZ_ModelIndep}, 
giving consistent results to the model-dependent approach~\cite{ref:belle_GGSZ}. The LHCb collaboration
has also released results of a model-independent GGSZ analysis~\cite{ref:lhcb_GGSZ}.

In order to determine \g with the GLW method, the analyses measure the direct \CP-violating 
partial decay rate asymmetries
\begin{eqnarray}
\label{eq:ACPpm}
A_{\CP\pm}^{(*)} & \equiv & \frac{\Gamma(\Bm \to \DDstar_{\CP\pm}\Km)-\Gamma(\Bp \to \DDstar_{\CP\pm}\Kp)}
                              {\Gamma(\Bm \to \DDstar_{\CP\pm}\Km)+\Gamma(\Bp \to \DDstar_{\CP\pm}\Kp)},\nn \\
\end{eqnarray}
and the ratios of charge-averaged partial rates using \D decays to \CP and flavor eigenstates,
\begin{eqnarray}
\label{eq:RCPpm}
R_{\CP\pm}^{(*)} & \equiv & 2\frac{\Gamma(\Bm\to \DDstar_{\CP\pm}\Km)+\Gamma(\Bp\to \DDstar_{\CP\pm}\Kp)}
                              {\Gamma(\Bm\to \DDstarz \Km)+\Gamma(\Bp\to \DDstarzb \Kp)},\nn\\
\end{eqnarray}
where $\DDstar_{\CP\pm}$ refers to the \CP eigenstates of the \DDstar meson system.
We select \D mesons in the \CP-even eigenstates $\pim\pip$ and $\Km\Kp$ ($\D_{\CP+}$),
in the \CP-odd eigenstates $\KS\piz$, $\KS\phi$, and $\KS\omega$ ($\D_{\CP-}$), and in the non-\CP eigenstate
$\Km\pip$ (\Dz from $\Bm\to\Dz \hm$) or $\Kp\pim$ (\Dzb from $\Bp\to\Dzb \hp$).
We recontruct \Dstar mesons in the states $\D\piz$ and $\D\gamma$.
The observables $A_{\CP\pm}^s$ and $R_{\CP\pm}^s$ for $\Bpm\to\D\Kstarpm$ decays are defined similarly.

For later convenience, the GLW observables can be related to \zbzbstpm and \zspm (neglecting mixing and \CP violation in neutral \D decays) as
\begin{eqnarray}
\label{eq:ACPpm2}
A_{\CP\pm}^{(*)}  &=& \pm \frac{\xbxbstm - \xbxbstp} {1 + |\zbzbst|^2 \pm (\xbxbstm + \xbxbstp) },
\end{eqnarray}
and
\begin{eqnarray}
\label{eq:RCPpm2}
R_{\CP\pm}^{(*)} &=& 1 + |\zbzbst|^2 \pm (\xbxbstm + \xbxbstp), 
\end{eqnarray}
where $|\zbzbst|^2$ is the average value 
of 
$|\zbzbstp|^2$ and $|\zbzbstm|^2$.
For $\Bpm\to\D\Kstarpm$ decays, similar relations to Eqs.~(\ref{eq:ACPpm2}) and~(\ref{eq:RCPpm2}) hold, 
with $\kappa=1$, 
since the effects 
of the non-\Kstar $\B\to\D\K\pi$ events and the width of the \Kstar are incorporated into the systematic uncertainties
of the $A_{\CP\pm}^s$ and $R_{\CP\pm}^s$ measurements~\cite{ref:GLWADS_d0kstar}.

Table~\ref{tab:GLWresults} summarizes the results obtained for the GLW observables.
In order to avoid overlaps with the samples selected in the Dalitz plot analysis, the results
for $\Bpm\to \D_{\CP-} \Kpm$ decays are corrected removing the contribution from $\D_{\CP-} \to \KS\phi$, $\phi\to K^+K^-$ candidates~\cite{ref:GLW_d0k}.
For the decays $\Bpm\to \Dstar_{\CP-}[\D_{\CP-}\piz] \Kpm$, 
$\Bpm\to \Dstar_{\CP+}[\D_{\CP-}\gamma] \Kpm$, and $\Bpm\to \D_{\CP-} \Kstarpm$, such information is not available.
In this case, the overlap is accounted for
by increasing the 
uncertainties quoted in Refs.~\cite{ref:GLW_dstar0k,ref:GLWADS_d0kstar}
by 10\% while keeping the central values unchanged. The 10\% increase in the experimental 
uncertainties is approximately the change observed in $\Bpm\to \D_{\CP-} \Kpm$ decays
when excluding or including $D\to \KS\phi$ in the measurement.
The impact on the combination has been found to be negligible. 

\begin{table}[!htb]
\caption{\label{tab:GLWresults}
GLW observables measured for the $\Bpm\to\D\Kpm$ (based on 467 million \BB pairs)~\cite{ref:GLW_d0k}, 
$\Bpm\to\Dstar\Kpm$ (383 million \BB pairs)~\cite{ref:GLW_dstar0k}, 
and $\Bpm\to\D\Kstarpm$ (379 million \BB pairs)~\cite{ref:GLWADS_d0kstar} decays, corrected 
removing the contribution from $\D_{\CP-} \to \KS\phi$, $\phi\to K^+K^-$ candidates.
The first uncertainty is statistical, the second is systematic.
}
\begin{center}
\begin{ruledtabular}
\begin{tabular}{ lrr }
            &   \CP-even \phzz\phz    &  \CP-odd \phzz\phz \\ \hline
$R_{\CP\pm}$   &  $1.18\pm 0.09\pm 0.05$     &  $1.03\pm 0.09\pm 0.04$ \\ 
$A_{\CP\pm}$   &  $0.25\pm 0.06\pm 0.02$     &  $-0.08\pm 0.07\pm 0.02$ \\ 
$R_{\CP\pm}^*$ &  $1.31\pm 0.13\pm 0.04$     &  $1.10\pm 0.13\pm 0.04$ \\ 
$A_{\CP\pm}^*$ &  $-0.11\pm 0.09\pm 0.01$     &  $0.06\pm 0.11\pm 0.02$ \\ 
$R_{\CP\pm}^s$ &  $2.17\pm0.35\pm0.09$ & $1.03\pm0.30\pm0.14$ \\ 
$A_{\CP\pm}^s$ &  $0.09\pm0.13\pm0.06$  & $-0.23\pm0.23\pm0.08$ \\ 
\end{tabular}
\end{ruledtabular}
\end{center}
\end{table}

As in the case of the GGSZ observables,
$A_{\CP\pm}^{(*)}$, $A_{\CP\pm}^s$, $R_{\CP\pm}^{(*)}$, and $R_{\CP\pm}^s$ have 
Gaussian uncertainties near the best solution, with small statistical and systematic correlations, as given in Ref.~\cite{ref:GLW_d0k}
for $\Bpm\to \D_{\CP-} \Kpm$ decays.
The GLW method has also been exploited by
the Belle~\cite{ref:belle_glw}, CDF~\cite{ref:cdf_glw},
and LHCb collaborations~\cite{ref:lhcb_glw}, with consistent results.

In the ADS method, the \Dz meson from the favored $\b \to \c \ubar \s$ amplitude is reconstructed 
in the doubly-Cabibbo-suppressed decay $\Kp\pim$, while the \Dzb from the $\b\to \u \cbar \s$ suppressed amplitude is
reconstructed in the favored decay $\Kp\pim$~\cite{ref:ADS_d0k_dstar0k_kpi,ref:GLWADS_d0kstar}. The product branching fractions for these final 
states, which we denote as $\Bm \to [\Kp\pim]_\D \Km$, $\Bm \to [\Kp\pim]_\Dstar \Km$, $\Bm \to [\Kp\pim]_\D \Kstarm$, and their \CP conjugates,
are small ($\sim 10^{-7}$). However, the two interfering amplitudes are of the same order
of magnitude, allowing for possible large \CP asymmetries. 
We measure charge-specific ratios for \Bp and \Bm decay rates to the ADS final states, which are defined as
\begin{eqnarray}
  \label{eq:Rpm_ADS}
  R_\pm^{(*)} & \equiv & \frac{ \Gamma(\Bpm \to [\Kmp\pipm]_{\D}^{(*)} \Kpm) } { \Gamma(\Bpm \to [\Kpm\pimp]_{\D}^{(*)} \Kpm) },
\end{eqnarray}
and similarly for $R_\pm^s$,
where the favored decays 
$\Bm \to [\Km\pip]_\D \Km$, $\Bm \to [\Km\pip]_\Dstar \Km$, and $\Bm \to [\Km\pip]_\D \Kstarm$
serve as normalization so that many systematic uncertainties cancel.
The rates in Eq.~(\ref{eq:Rpm_ADS}) depend on \g and the \B decay hadronic parameters. They are related to \zbzbstpm and \zspm through
\begin{eqnarray}
  \label{eq:Rpm_ADS_vs_xy}
R_\pm^{(*)} & = & {\rbrbstpm}^2 + \rd^2 + 2 \lambda \rd \left[ \xbxbstpm \cos\deltad - \ybybstpm \sin\deltad \right],\nn\\
\end{eqnarray}
where $\rd = |{\cal A}(\Dz\to \Kp\pim)/{\cal A}(\Dz\to \Km\pip)|$ and \deltad
are the ratio between magnitudes of the suppressed and favored \D decay amplitudes and their relative strong phase, respectively.
As in Eq.~(\ref{eq:GGSZrates}), the symbol $\lambda$ for $\Bpm\to\Dstar\Kpm$ decays accounts for the different \CP parity of 
$\Dstar\to\D\piz$ and $\Dstar\to\D\gamma$.
The values of 
\rd and \deltad
are taken as external constraints in our analysis. 
As for the GLW method, the effects of other $\Bpm\to\D\KS\pipm$ events, not going through $\Kstarpm$, and the \Kstarpm width, 
are incorporated in the systematic uncertainties on $R_\pm^s$. Thus, similar relations hold for these
observables with $\kappa=1$.

The choice of the observables $R_\pm$ (and similarly for $R_\pm^{*}$ and $R_\pm^s$) rather than the original
ADS observables 
$R_{\mathrm{ADS}} \equiv (R_+ + R_-)/2$ and
$A_{\mathrm{ADS}} \equiv (R_- - R_+)/2R_{\mathrm{ADS}}$~\cite{ref:ads_theo} is motivated by the fact that
the set of variables 
$( R_{\mathrm{ADS}},A_{\mathrm{ADS}})$ is not well behaved since the uncertainty
on $A_{\mathrm{ADS}}$ depends on the central value of $R_{\mathrm{ADS}}$, while
$R_+$ and $R_-$ are statistically independent observables.
Although systematic
uncertainties are largely correlated, the 
measurements of $R_+$ and $R_-$ are effectively uncorrelated since the total uncertainties are
dominated by the statistical component.

We have also reconstructed $\Bpm \to [\Kmp\pipm\piz]_\D \Kpm$ decays~\cite{ref:ADS_d0k_kpipi0} from which
the observables $R_\pm^{\K\pi\piz}$ have been measured, which are related to the GGSZ observables as
\begin{eqnarray}
  \label{eq:Rpm_ADSKpipiz_vs_xy}
R_\pm^{\K\pi\piz} & = & {\rbpm}^2 + \rKpipiz^2 + 2 \kappaKpipiz \rKpipiz \nn\\
      & &  \times \left[ \xbpm \cos\deltaKpipiz - \ybpm \sin\deltaKpipiz \right],
\end{eqnarray}
where \kappaKpipiz is a \D decay coherence factor similar to that defined in Eq.~(\ref{eq:kappa}) for the $\Bpm\to\D\KS\pipm$ decay,
and where \rKpipiz and \deltaKpipiz are hadronic parameters for $\Dz \to \Kpm\pimp\piz$ decays analogous to \rd and \deltad.

\begin{figure}[!htb]
\begin{center}
\begin{tabular} {cc}
  \epsfig{file=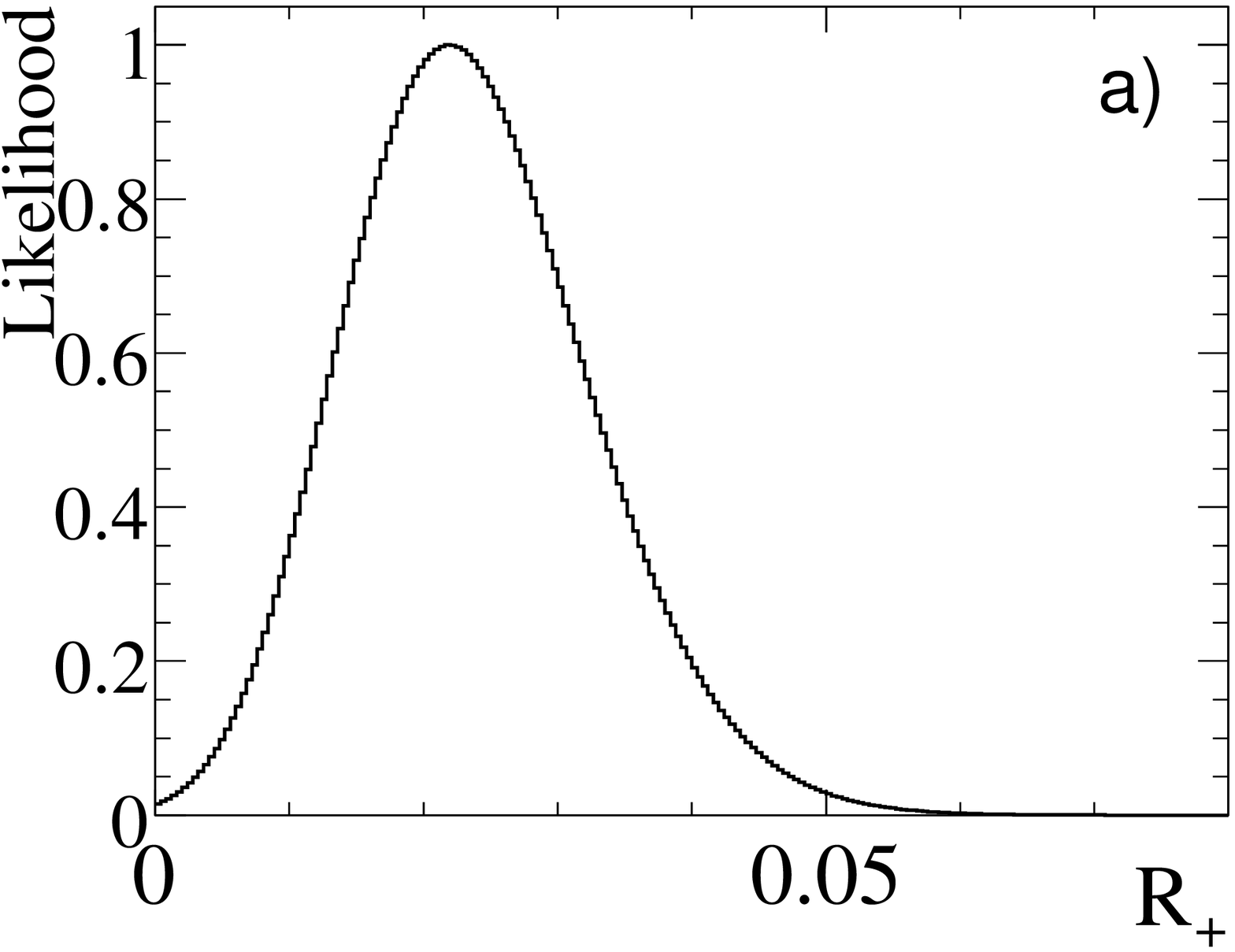,width=0.48\linewidth} &
  \epsfig{file=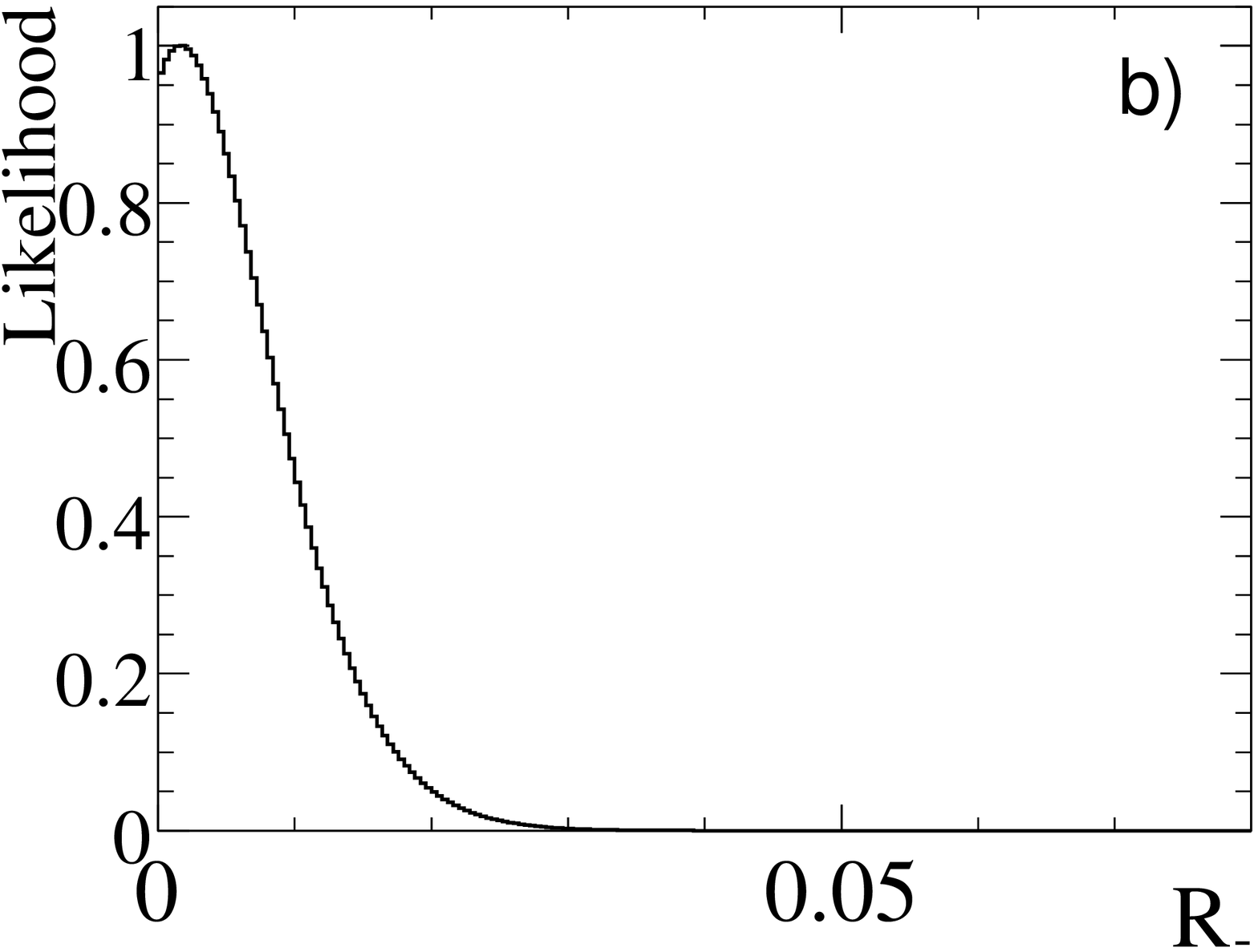,width=0.48\linewidth} \\
  \epsfig{file=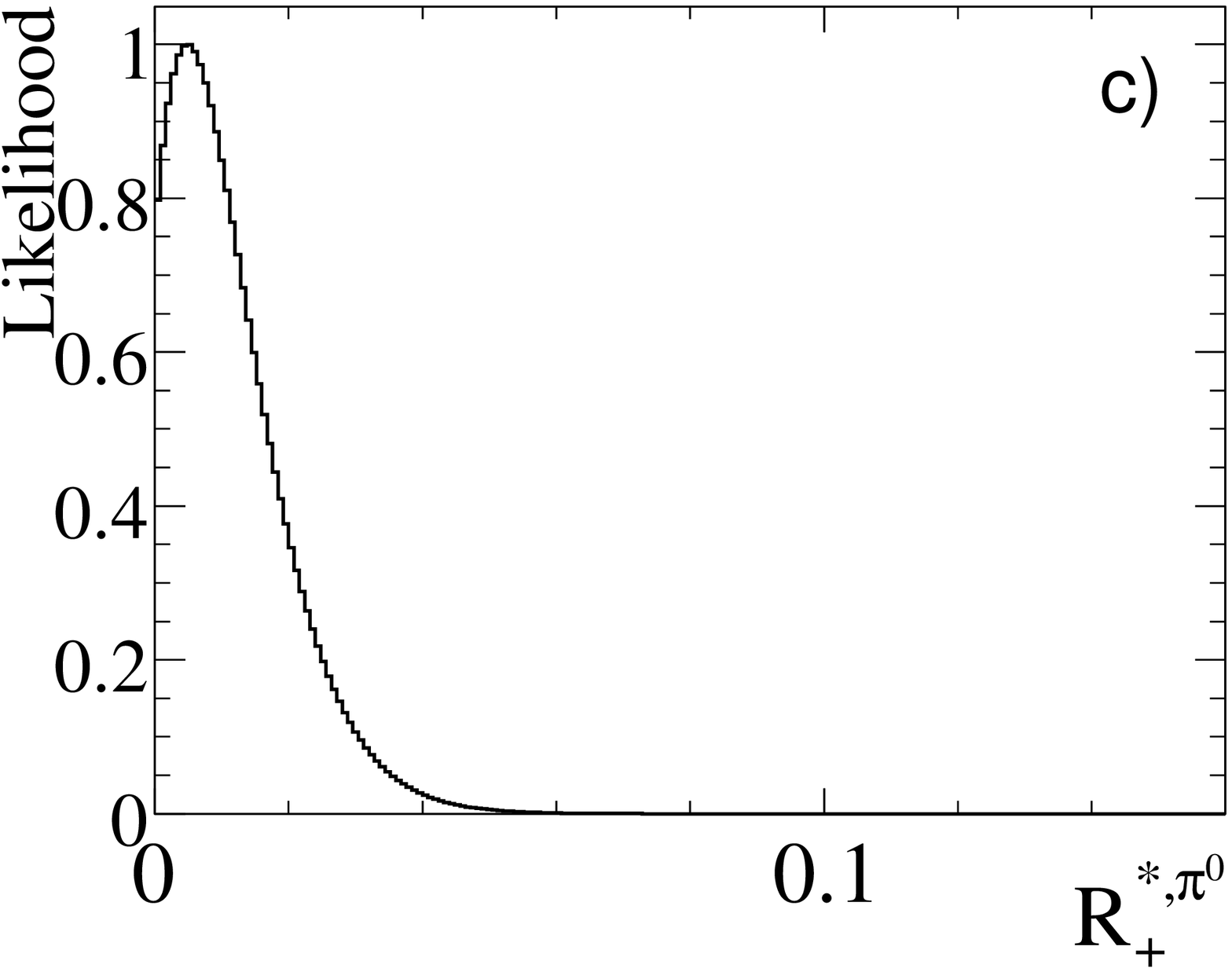,width=0.48\linewidth} &
  \epsfig{file=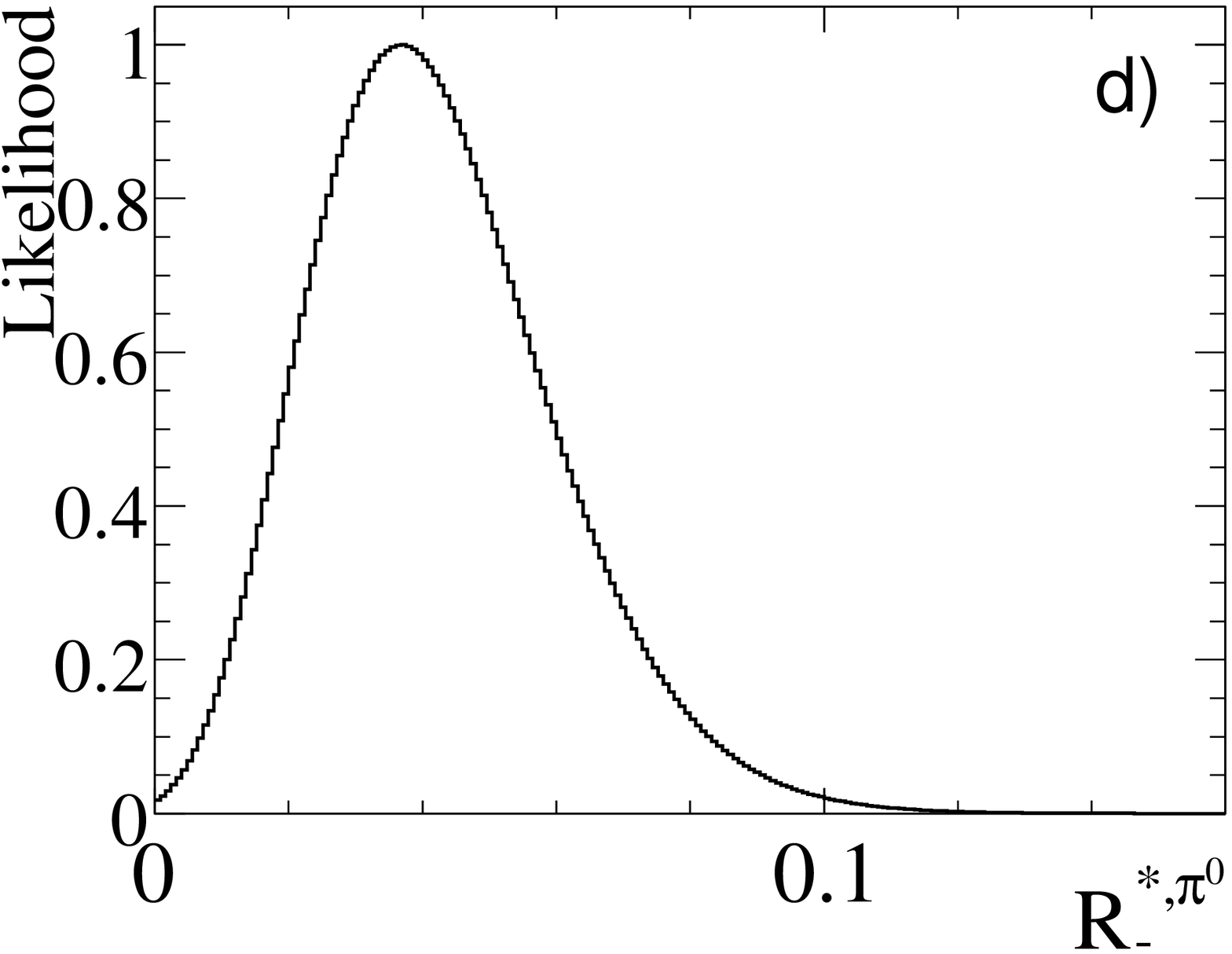,width=0.48\linewidth} \\
  \epsfig{file=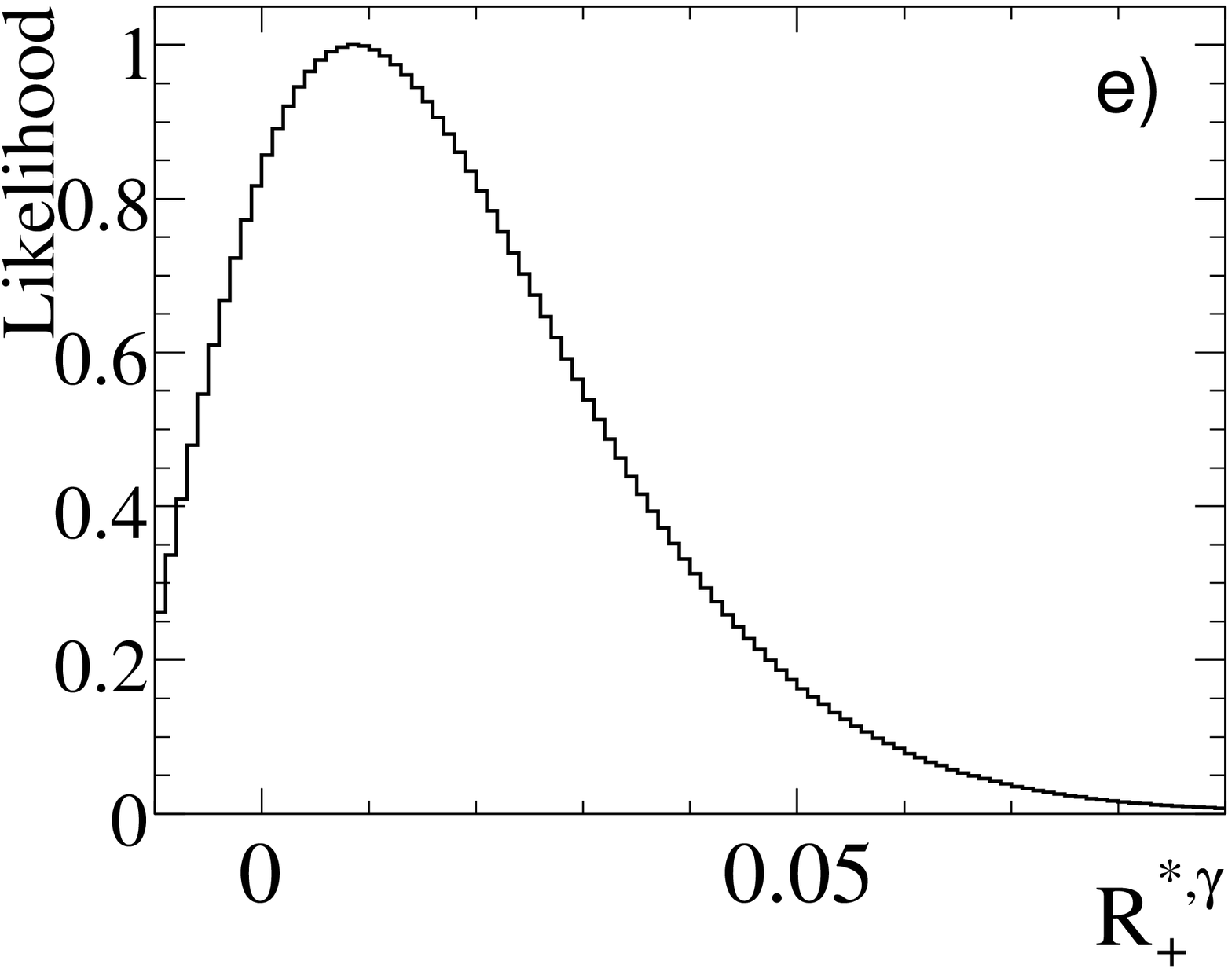,width=0.48\linewidth} &
  \epsfig{file=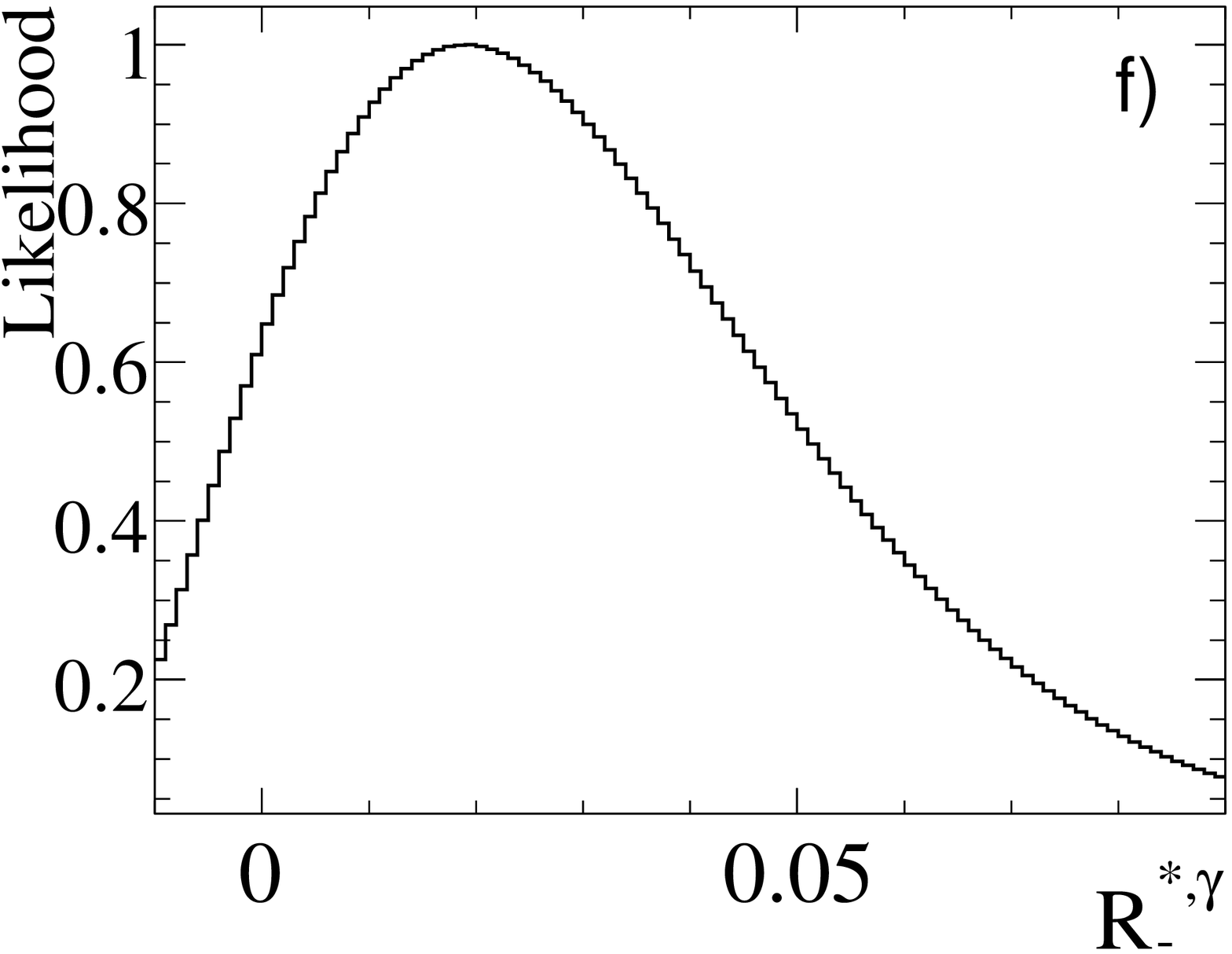,width=0.48\linewidth} \\
  \epsfig{file=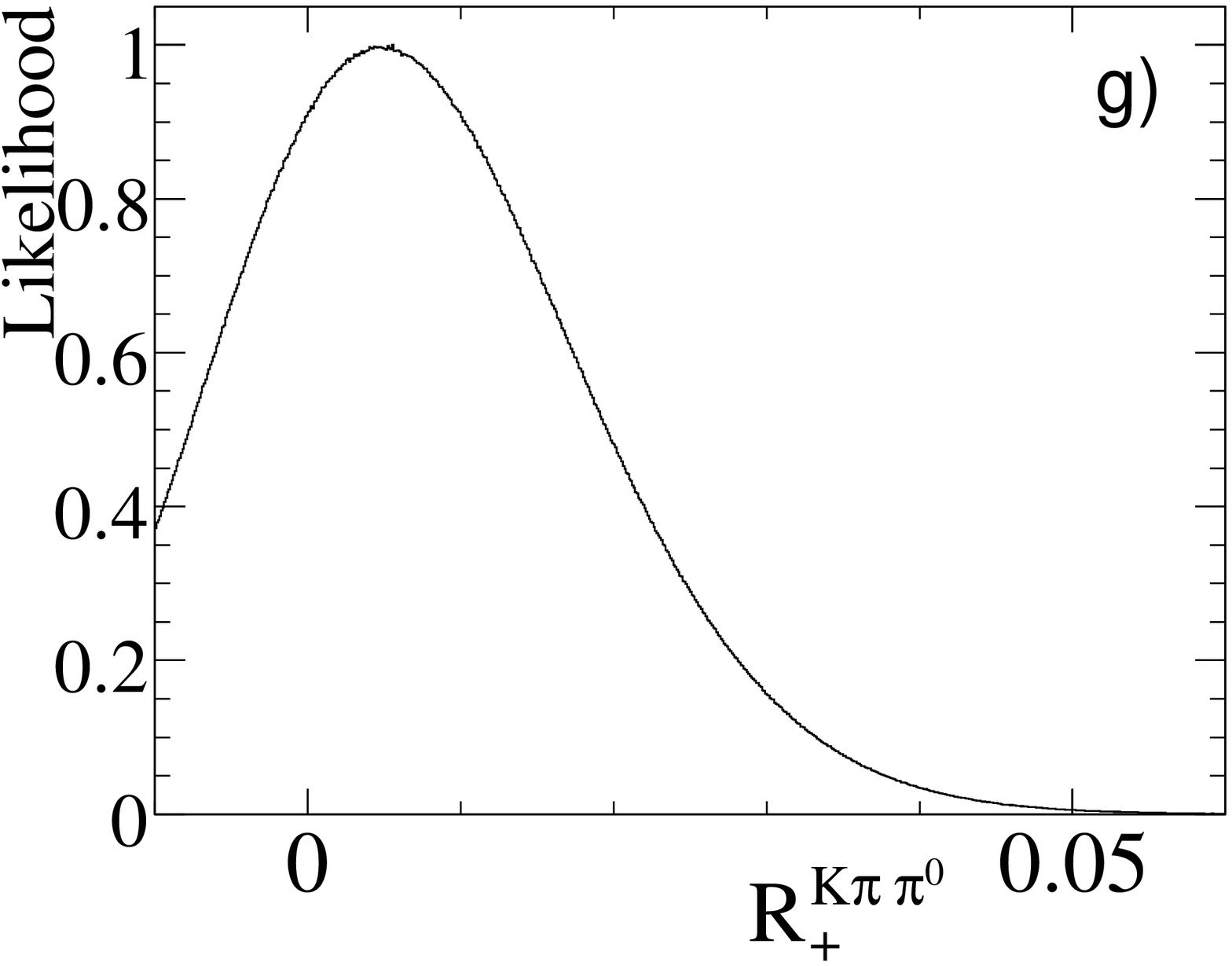,width=0.48\linewidth} &
  \epsfig{file=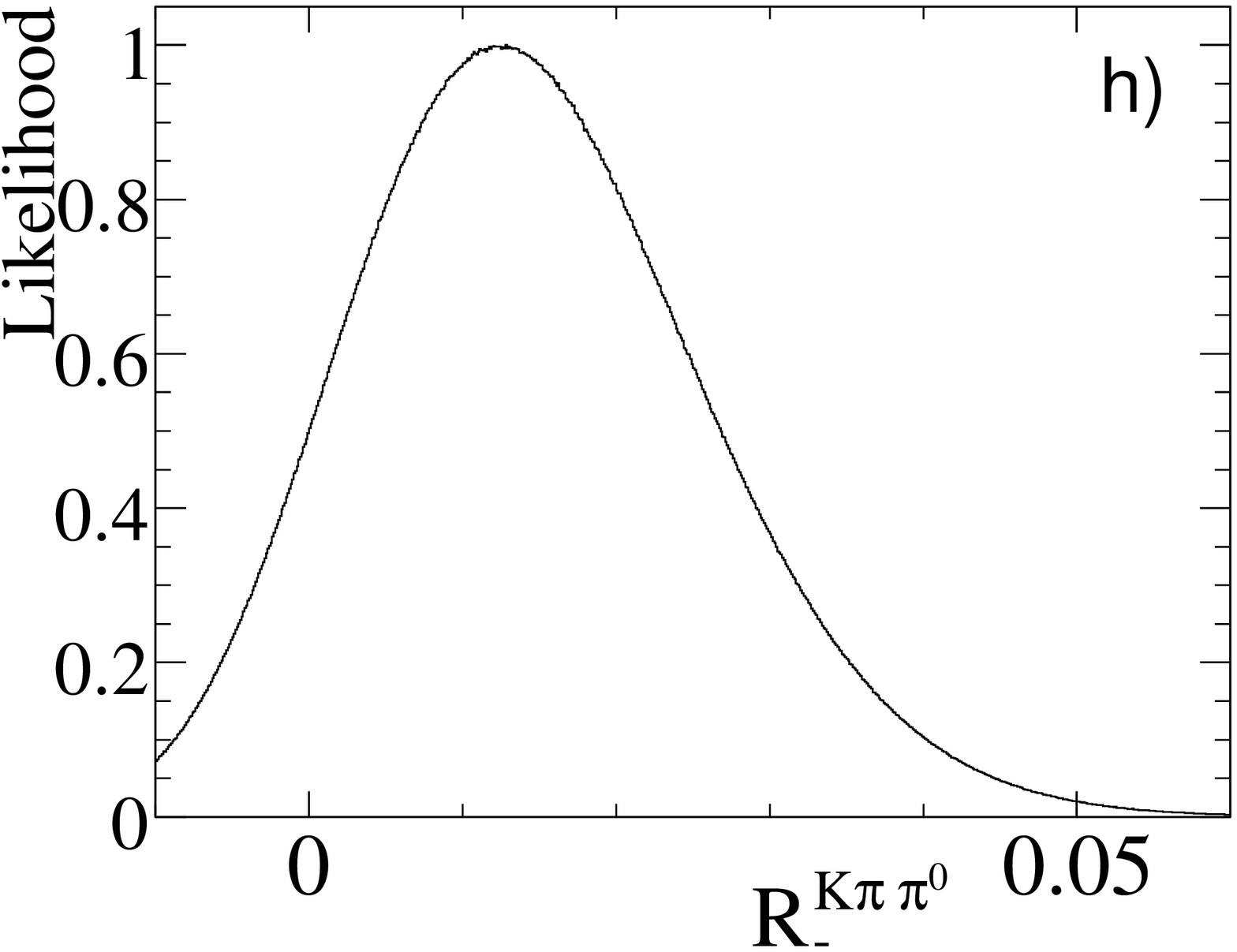,width=0.48\linewidth} \\
\end{tabular}
\caption{\label{fig:ADSlikelihood} 
Experimental likelihoods as functions of the ADS charge-specific ratios 
$R_\pm$ (a,b), $R_\pm^{*}$~$[\D\piz]$ (c,d), $R_\pm^{*}$~$[\D\gamma]$ (e,f), and $R_\pm^{\K\pi\piz}$ (g,h),
from Refs.~\cite{ref:ADS_d0k_dstar0k_kpi,ref:ADS_d0k_kpipi0}, including systematic uncertainties. The \pdfs are normalized so that their maximum values
are equal to $1$.
These distributions are well parameterized by sums of two asymmetric Gaussian functions with mean values as given in Table~\ref{tab:ADSresults}.
}
\end{center}
\end{figure}

Table~\ref{tab:ADSresults} summarizes the measurements of the ADS charge-specific ratios for the different final states.
Contrary to the case of the GGSZ and GLW observables, $R_\pm^{(*)}$, $R_\pm^s$, and $R_\pm^{\K\pi\piz}$ do not have Gaussian behavior.
The experimental likelihood function for each of the four decay modes, shown in Fig.~\ref{fig:ADSlikelihood} for 
$\Bpm\to\D\Kpm$ and $\Bpm\to\Dstar\Kpm$ decays,
is well described around the best solution by an 
analytical \pdf composed of the sum of two asymmetric Gaussian functions. 
For the $\Bpm\to\D\Kstarpm$ channel, we use instead a simple Gaussian approximation since in this case
the experimental likelihood scans are not available. The effect of this approximation has been verified to be negligible, 
given the small statistical weight of this sample in the combination.  
Measurements using the ADS technique have also been performed by the Belle~\cite{ref:belle_ads,ref:belle_ads_DKstar}, 
CDF~\cite{ref:cdf_ads}, and LHCb collaborations~\cite{ref:lhcb_glw}, with consistent results.

\begin{table}[!htb]
\caption{ADS observables included into the combination for  $\Bpm\to\D\Kpm$ with $\D\to K\pi$ (based on 467 million \BB pairs) and
$\D\to K\pi\piz$  (based on 474 million \BB pairs), $\Bpm\to\Dstar\Kpm$ (467 million \BB pairs), 
and $\Bpm\to\D\Kstarpm$ (379 million \BB pairs) decays~\cite{ref:ADS_d0k_dstar0k_kpi,ref:GLWADS_d0kstar,ref:ADS_d0k_kpipi0}. 
The first uncertainty is statistical, the second is systematic.}
\label{tab:ADSresults}
\begin{center}
\begin{ruledtabular}
\begin{tabular}{lrr}
                   &  \Bp \phzz\phzz\phzz            & \Bm \phzz\phzz\phzz  \\ \hline
$R_{\pm}$           & $ 0.022\pm0.009\pm 0.003$   & $ 0.002\pm 0.006 \pm 0.002$\\
$R_{\pm}^{*}~[\D\piz]$   & $  0.005\pm 0.008\pm 0.003$ & $ 0.037\pm 0.018 \pm 0.009$\\
$R_{\pm}^{*}~[\D\gamma]$  & $ 0.009\pm 0.016 \pm 0.007$ & $0.019\pm 0.023 \pm 0.012$ \\
$R_{\pm}^s$          & $ 0.076\pm 0.042 \pm 0.011$ & $0.054\pm 0.049 \pm 0.011$ \\
$R_{\pm}^{\K\pi\piz}$   & $ 0.005\phantom{.}^{+0.012}_{-0.010}\phantom{.}^{+0.001}_{-0.004}$\phantom{...}  & $0.012\phantom{.}^{+0.012}_{-0.010}\phantom{.}^{+0.002}_{-0.004}$\phantom{...} \\
\end{tabular}
\end{ruledtabular}
\end{center}
\end{table}

\section{ \bf Other Measurements}
Similar analyses related to \g measurement have been carried out
using the decay $\Bm \to \D \Km$ with the $\D \to \pip \pim \piz$ final state~\cite{ref:GGSZ_pipipi0},
and the neutral \B decay $\Bzb \to \D \Kstarb(892)^0$, $\Kstarb(892)^0 \to \Km \pip$, 
with $\D \to \KS\pip\pim$~\cite{ref:GGSZ_d0kstar0} 
and 
$\D \to \Kpm\pimp, \Kpm\pimp\piz, \Kpm\pimp\pipm\pimp$~\cite{ref:ADS_d0kstar0}.
For neutral \B decays, \rb is naively expected to be larger ($\approx 0.3$) 
because both interfering amplitudes are color suppressed and thus 
$c_F\approx 1$. However, the overall rate of events 
is smaller than for $\Bm \to \D \Kstarm$ decays.
The flavor of the neutral \B meson is tagged by the charge of the kaon
produced in the \Kstarzb decay, $\Kstarb(892)^0 \to \Km\pip$ or $\Kstar(892)^0 \to \Kp\pim$.

Experimental analyses of the time-dependent decay rates of $B\to \DDstarmp \pipm$ and $B\to \Dmp \rhopm$ decays
have also been used to constrain $\gamma$~\cite{ref:Dstpi_partialreco,ref:DpiDstpiDrho_fullreco}. 
In these decays, the interference occurs between the favored $b \to c \ubar d$ and
the suppressed $b\to u\cbar d$ tree amplitudes with and without $\Bz-\Bzb$ mixing,
resulting in a total weak phase difference $2\beta+\gamma$~\cite{ref:Dunietz1997}, where $\beta$ is the angle of the unitarity 
triangle defined as $\arg [-V_{cd}V_{cb}^*/V_{td}V_{tb}^*]$.
The magnitude ratios between the suppressed and favored amplitudes $r_{\DDstar\pi}$ and $r_{\D\rho}$
are expected to be $\approx 2\%$, 
and have to be estimated either by analyzing suppressed charged \B decays (e.g., $\Bp\to \Dp\piz$) with an isospin assumption or
from self-tagging neutral \B decays to charmed-strange mesons (e.g., $\Bz\to \Ds \pim$) assuming SU(3) flavor symmetry and
neglecting contributions from $W$-exchange diagrams~\cite{ref:Dunietz1997}.
Performing a time-dependent Dalitz plot analysis of $\B\to \Dmp \Kz\pipm$
decays~\cite{ref:B0toDKspi} could in principle avoid the problem of
the smallness of $r$. In these decays the two interfering amplitudes
are color suppressed, and \rb is expected to be $\approx 0.3$
but the overall rate of events is too small with the current data
sample. 

In both cases, the errors on the experimental measurements are too
large for a meaningful determination of \g, and have not been included
in the combined determination of \g reported in this paper. However, 
these decay channels might provide important information in future experiments.

\section{ \bf Combination Procedure}

We combine all the GGSZ, GLW, and ADS observables (34 in total) to extract \g in two different stages. 
First, we extract the best-fit values for the \CP-violating quantities \zbzbstpmtrue and \zspmtrue, 
whose definitions correspond to
those for the quantities \zbzbstpm and \zspm of the GGSZ analysis given in Eq.~(\ref{eq:cartesian}).

Their best-fit values are obtained by maximizing a
combined likelihood function constructed as the product of partial 
likelihood \pdfs for GGSZ, GLW, and ADS measurements.
The GGSZ likelihood function uses a 12-dimensional Gaussian
\pdf with measurements \zbzbstpm and \zspm and their covariance matrices for statistical, experimental, and amplitude model
uncertainties, and mean (expected) values \zbzbstpmtrue and \zspmtrue. Similarly, the GLW likelihood is formed as
the product of four-dimensional Gaussian \pdfs for each $B$ decay with measurements $A_{\CP\pm}^{(*)}$, $A_{\CP\pm}^s$, $R_{\CP\pm}^{(*)}$, $R_{\CP\pm}^s$
and their covariance matrices, and expected values given by Eqs.~(\ref{eq:ACPpm2}) and~(\ref{eq:RCPpm2}) after
replacing the \zbzbstpm and \zspm observables by the \zbzbstpmtrue and \zspmtrue parameters.
Finally, the ADS \pdf is built from the product of experimental likelihoods shown in Fig.~\ref{fig:ADSlikelihood}. 
With this construction, GGSZ, GLW, and ADS observables are taken as uncorrelated.
Similarly, the individual measurements are considered uncorrelated as the 
experimental uncertainties are dominated by the statistical component.

The combination requires external inputs for the \D hadronic parameters
\rd, \deltad, \rKpipiz, \deltaKpipiz, and \kappaKpipiz.
We assume Gaussian \pdfs for $\rd=0.0575\pm0.0007$~\cite{ref:hfag2012} and $\rKpipiz=0.0469\pm0.0011$~\cite{ref:pdg2010},
while for the other three we adopt asymmetric Gaussian
parameterizations based on the experimental likelihoods available either from world averages for $\deltad=(202.0^{+9.9}_{-11.2})^\circ$~\cite{ref:hfag2012} 
or from the CLEOc collaboration for $\deltaKpipiz=(47^{+14}_{-17})^\circ$ and $\kappaKpipiz=0.84\pm0.07$~\cite{ref:cleoKpipi0}.
The values of \deltad and \deltaKpipiz have been corrected for a shift of $180^\circ$ in the definition of 
the phases between Refs.~\cite{ref:ADS_d0k_dstar0k_kpi,ref:ADS_d0k_kpipi0} and Refs.~\cite{ref:hfag2012,ref:cleoKpipi0}.
The correlations between \rd and \deltad, and between \kappaKpipiz and \deltaKpipiz, 
are small and have been neglected. 
All five external observables are assumed to be
uncorrelated with the rest of the input observables. 

\begin{figure*}[!htb]
\begin{center}
\begin{tabular} {ccc}
  \epsfig{file=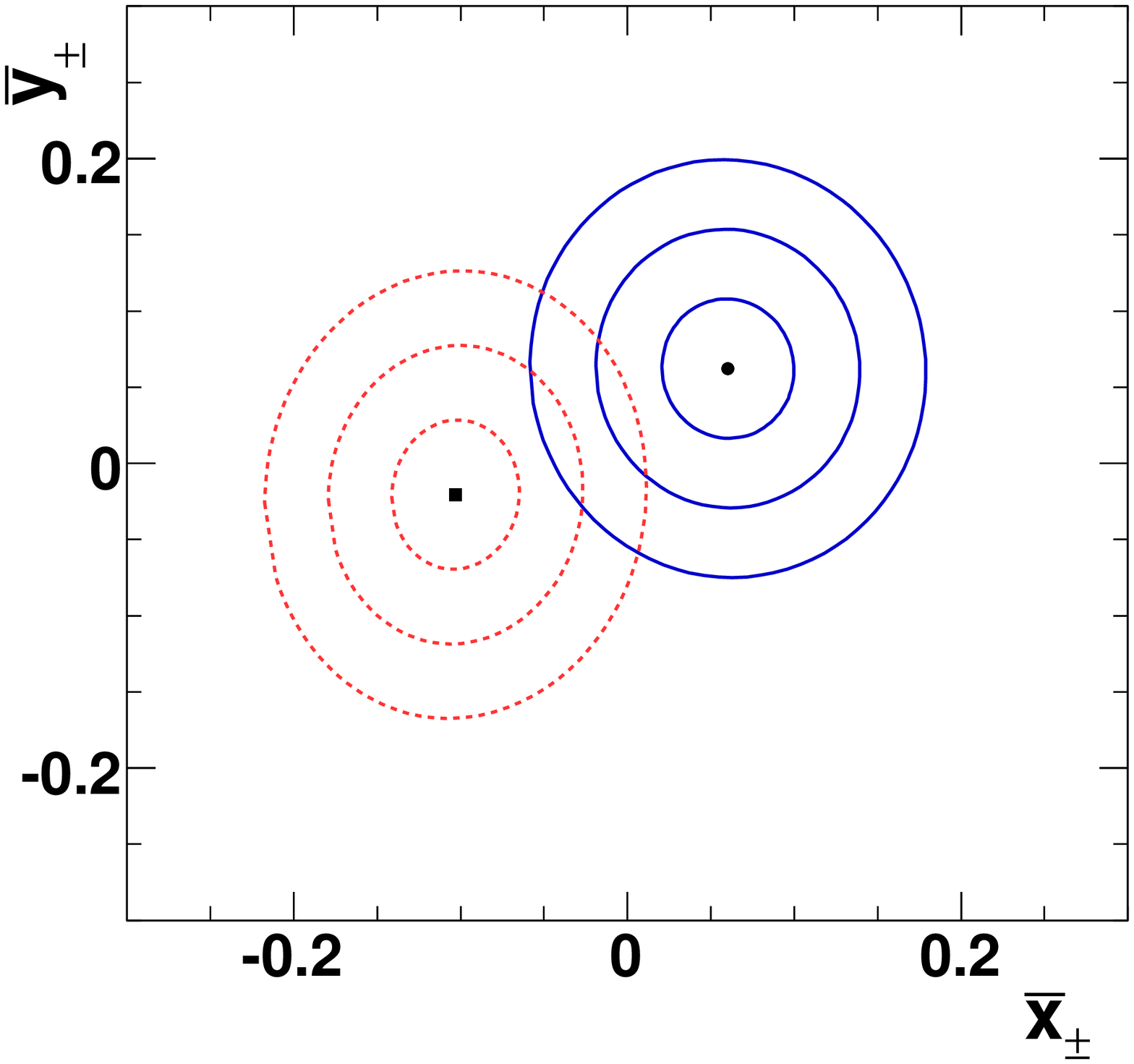,width=0.28\linewidth} &
  \epsfig{file=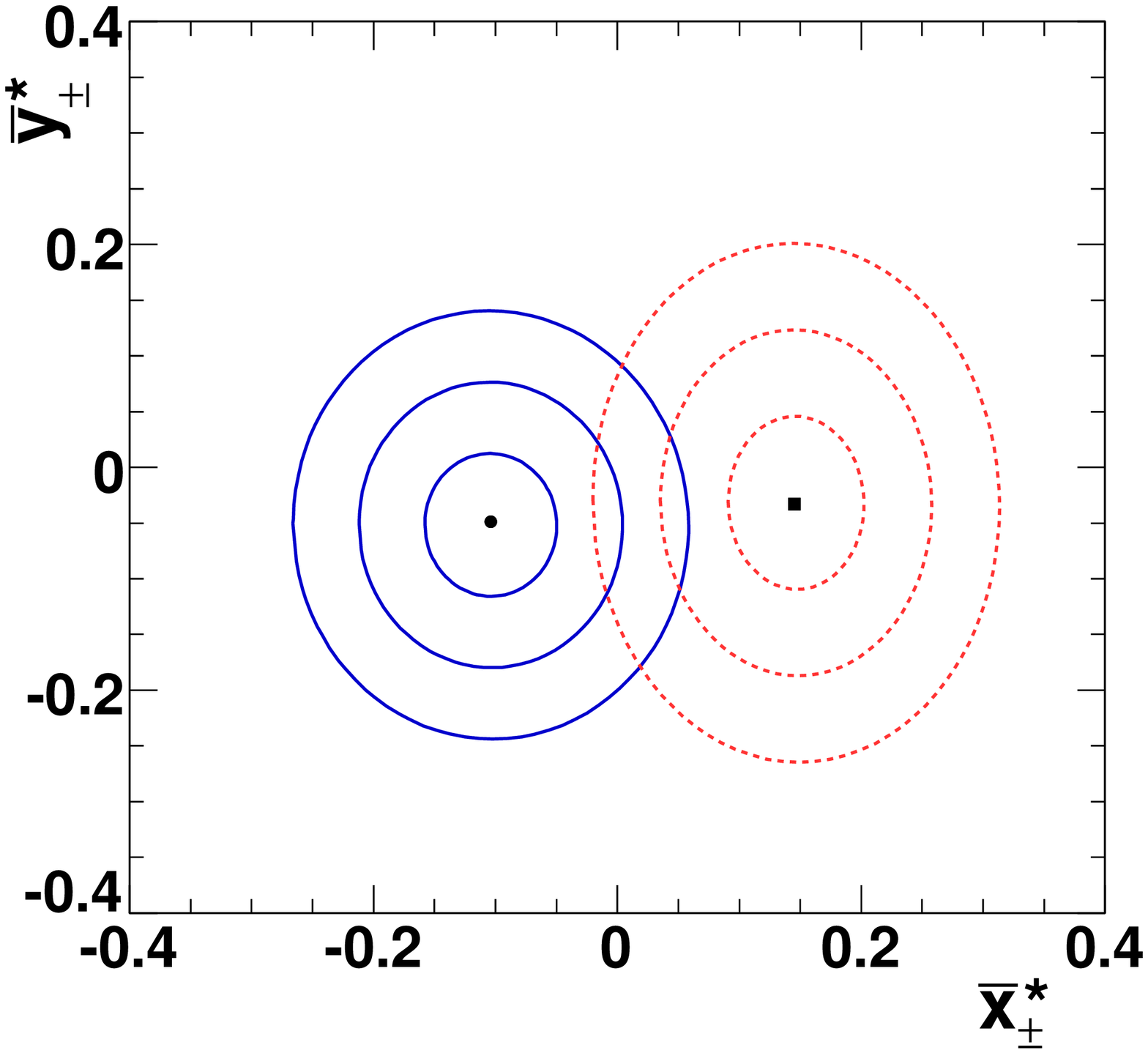,width=0.28\linewidth} &
  \epsfig{file=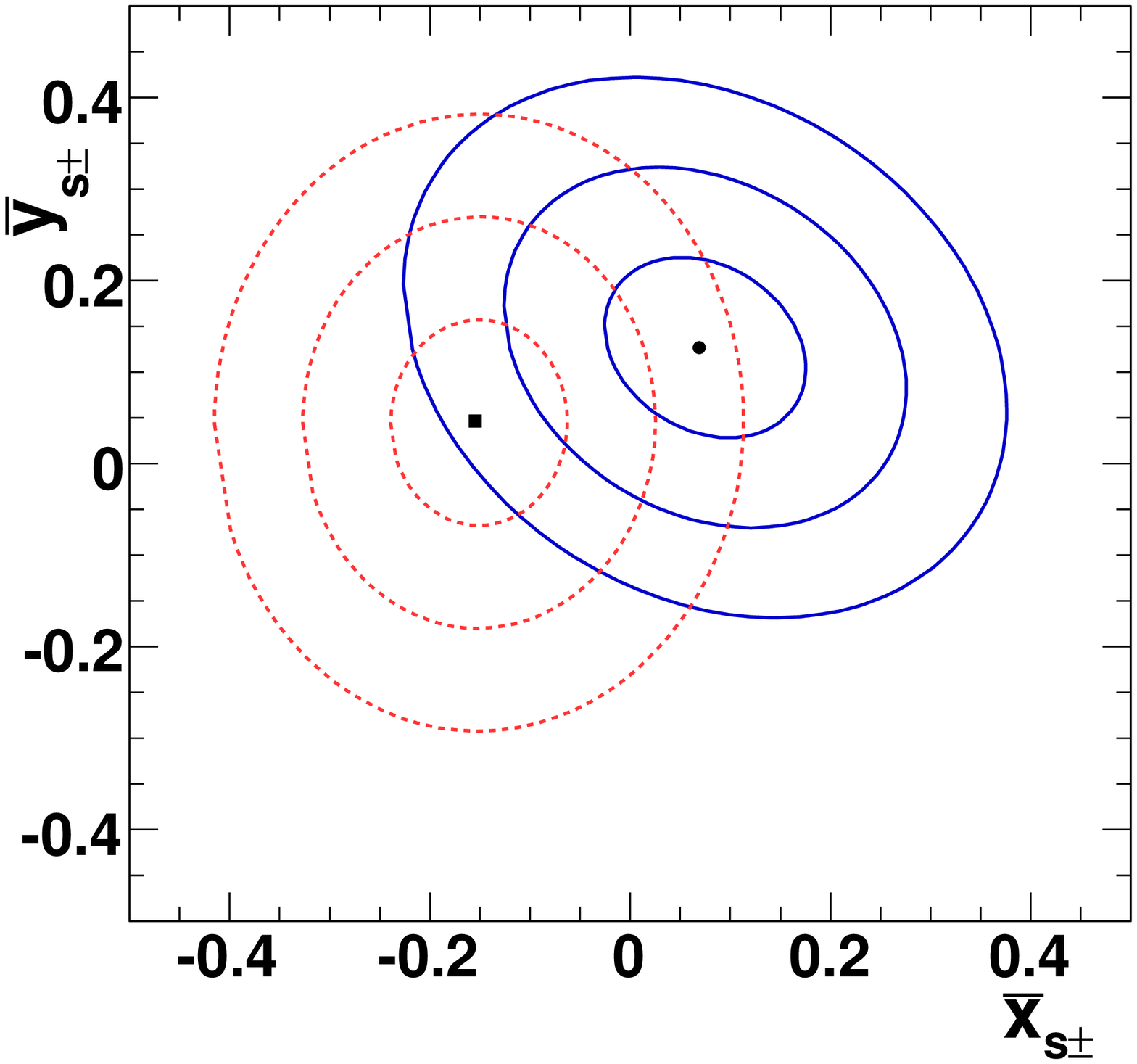,width=0.28\linewidth} \\
  \epsfig{file=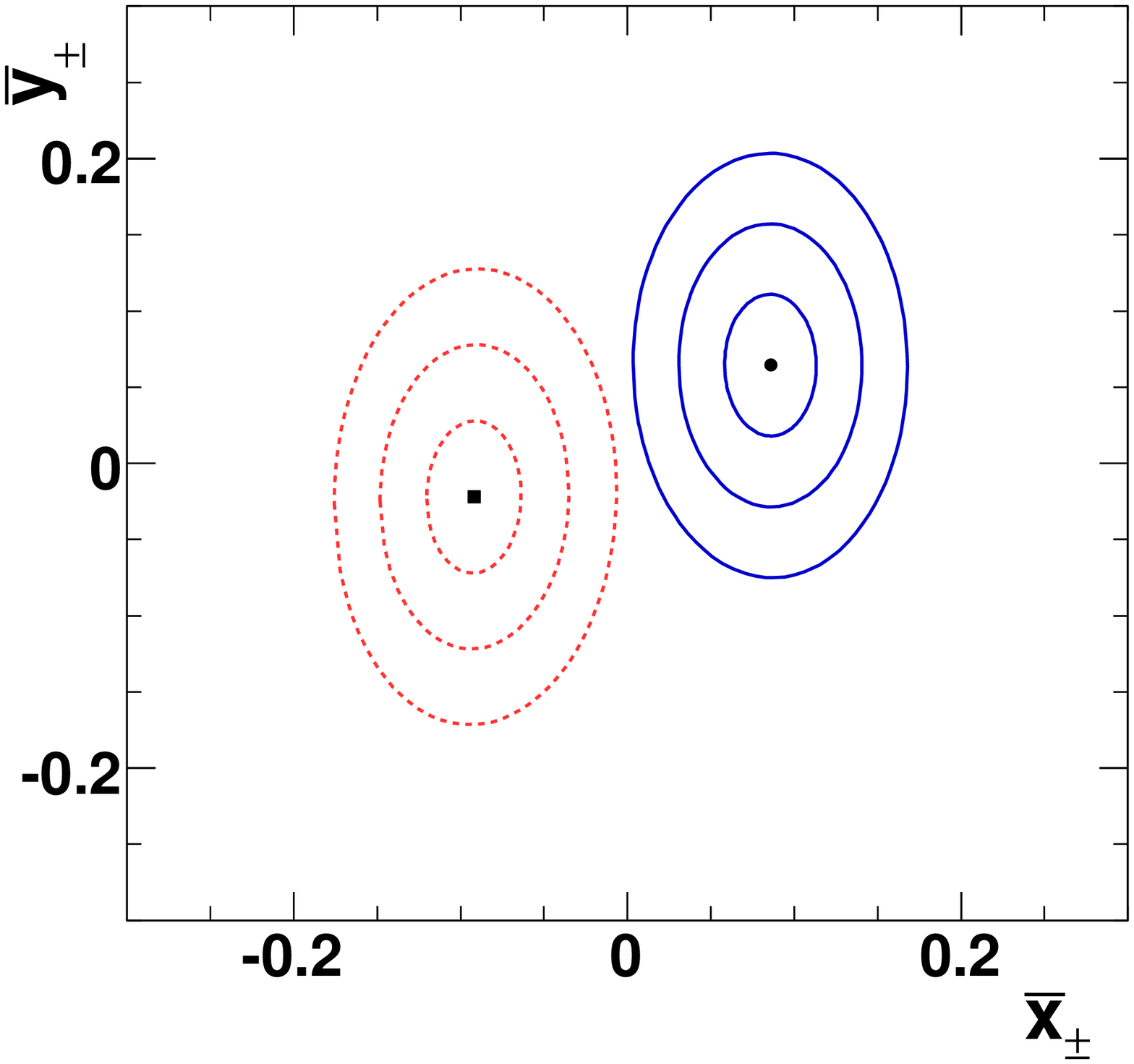,width=0.28\linewidth} &
  \epsfig{file=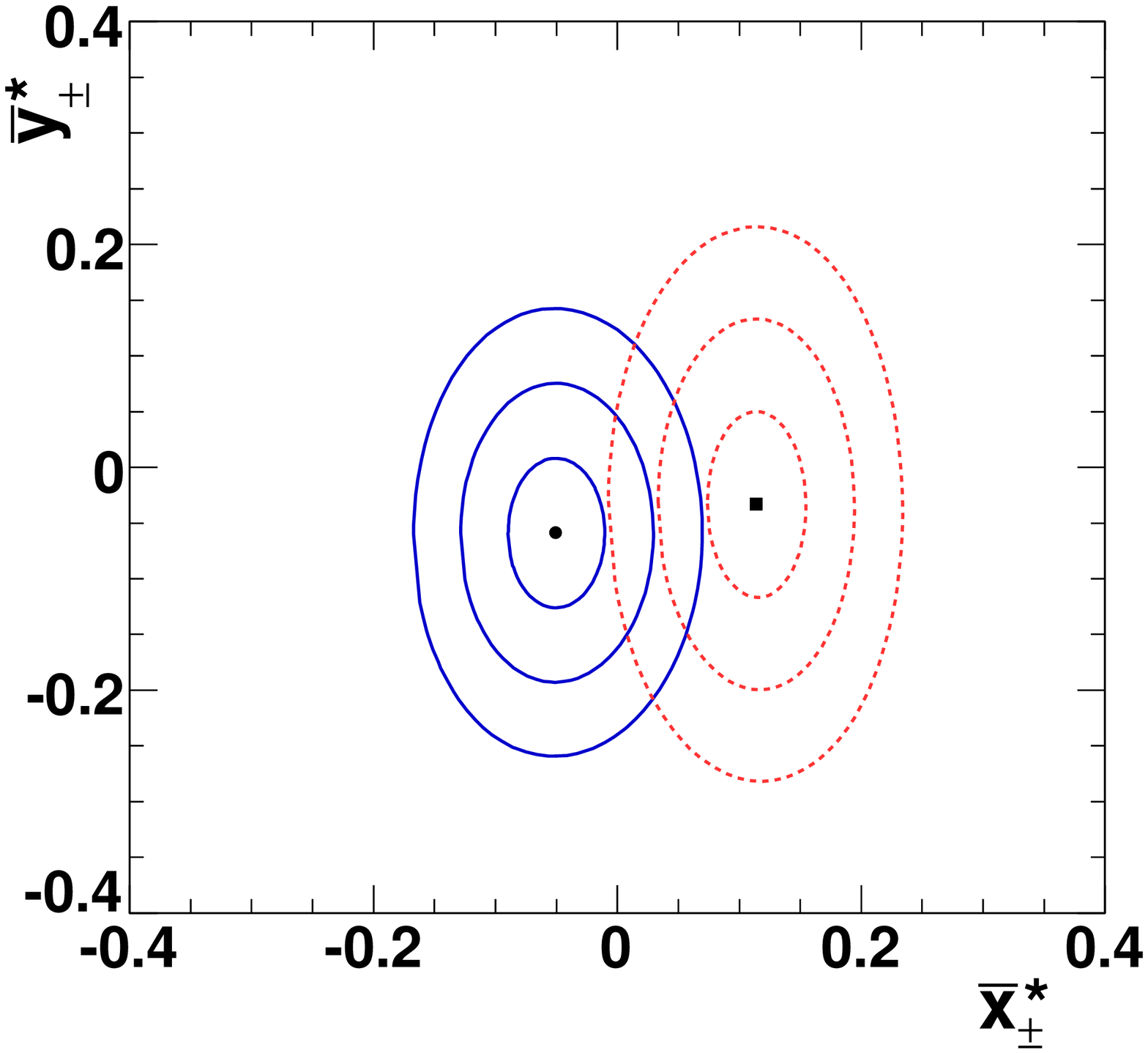,width=0.28\linewidth} &
  \epsfig{file=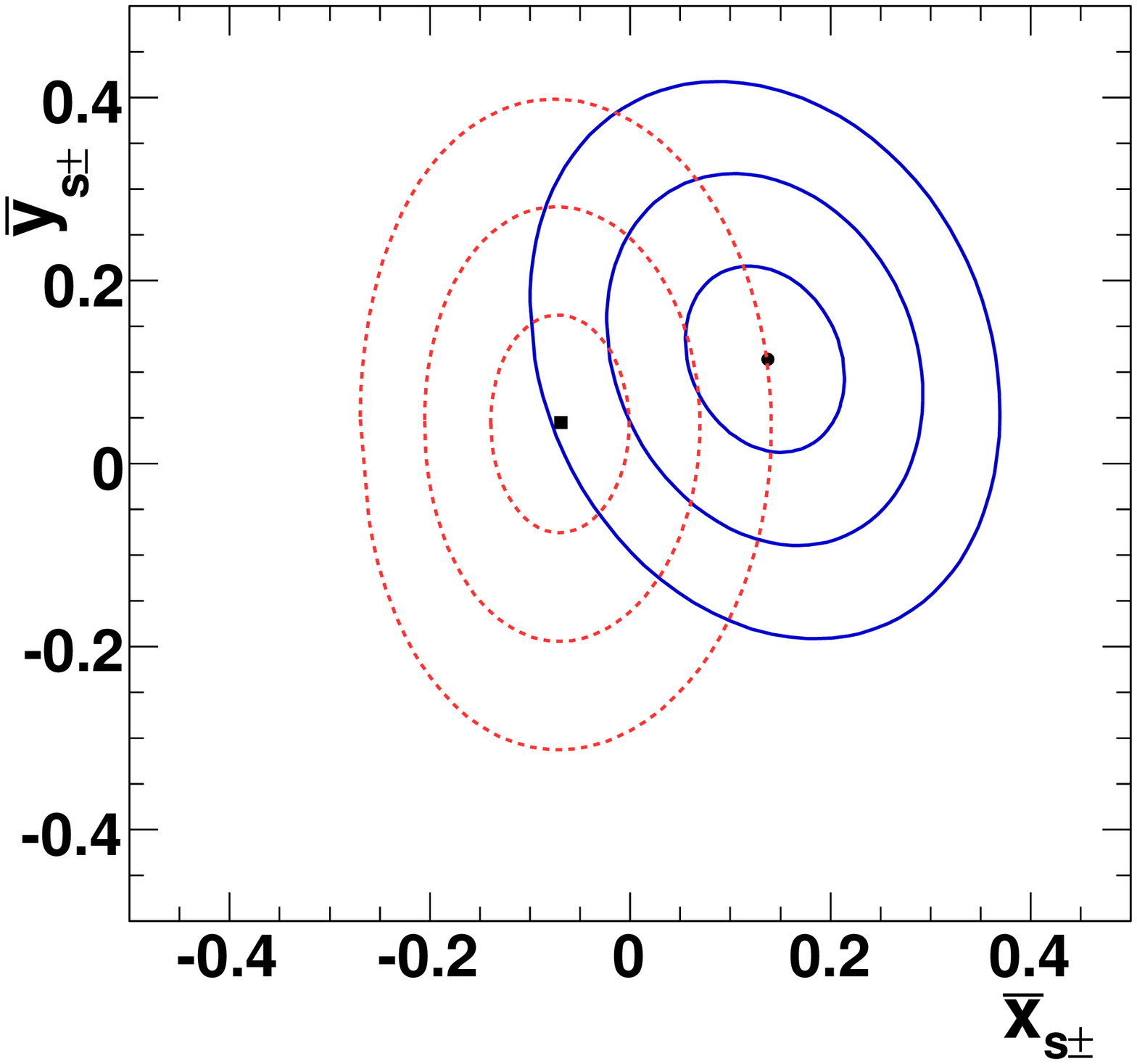,width=0.28\linewidth} \\
  \epsfig{file=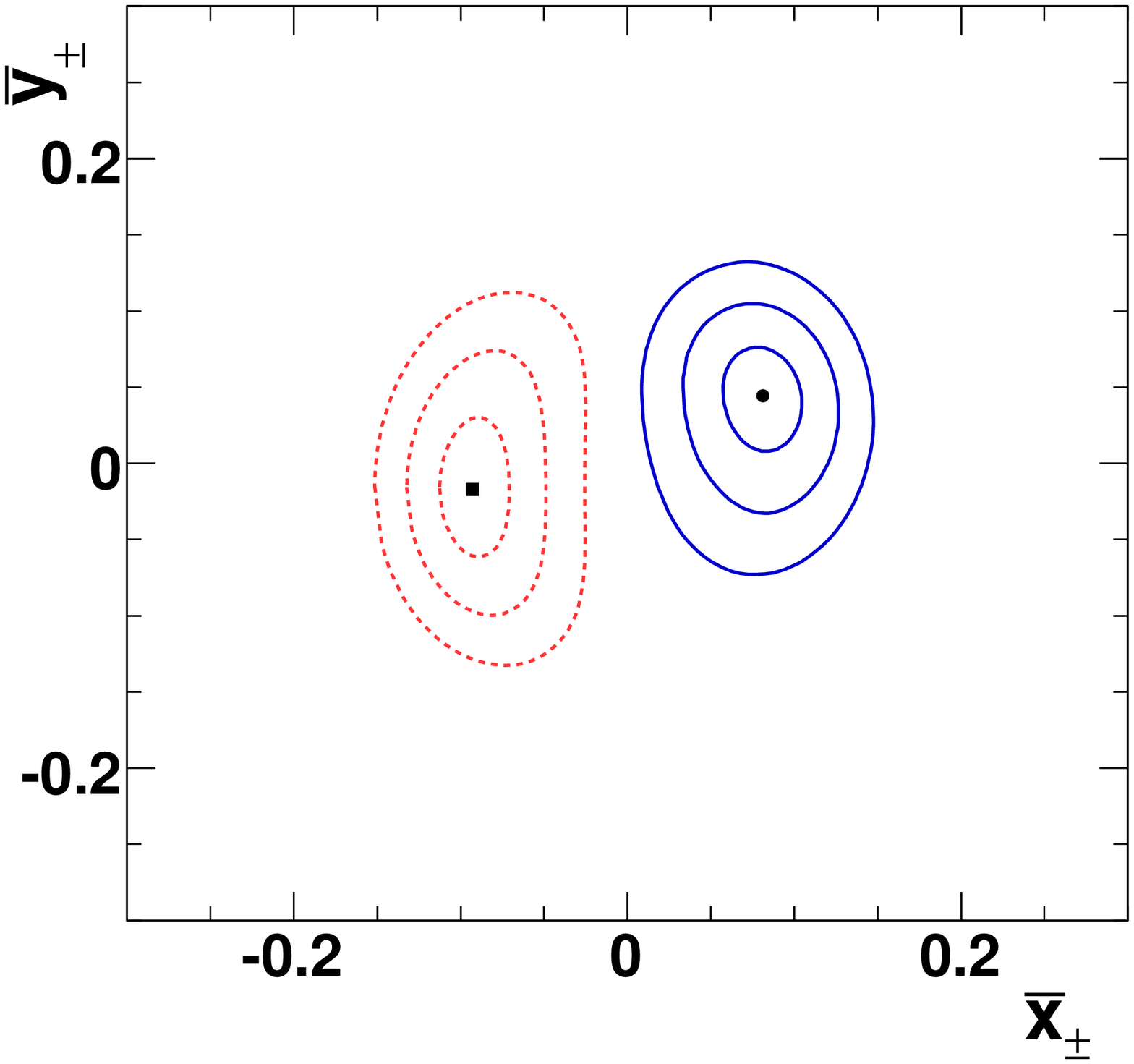,width=0.28\linewidth} &
  \epsfig{file=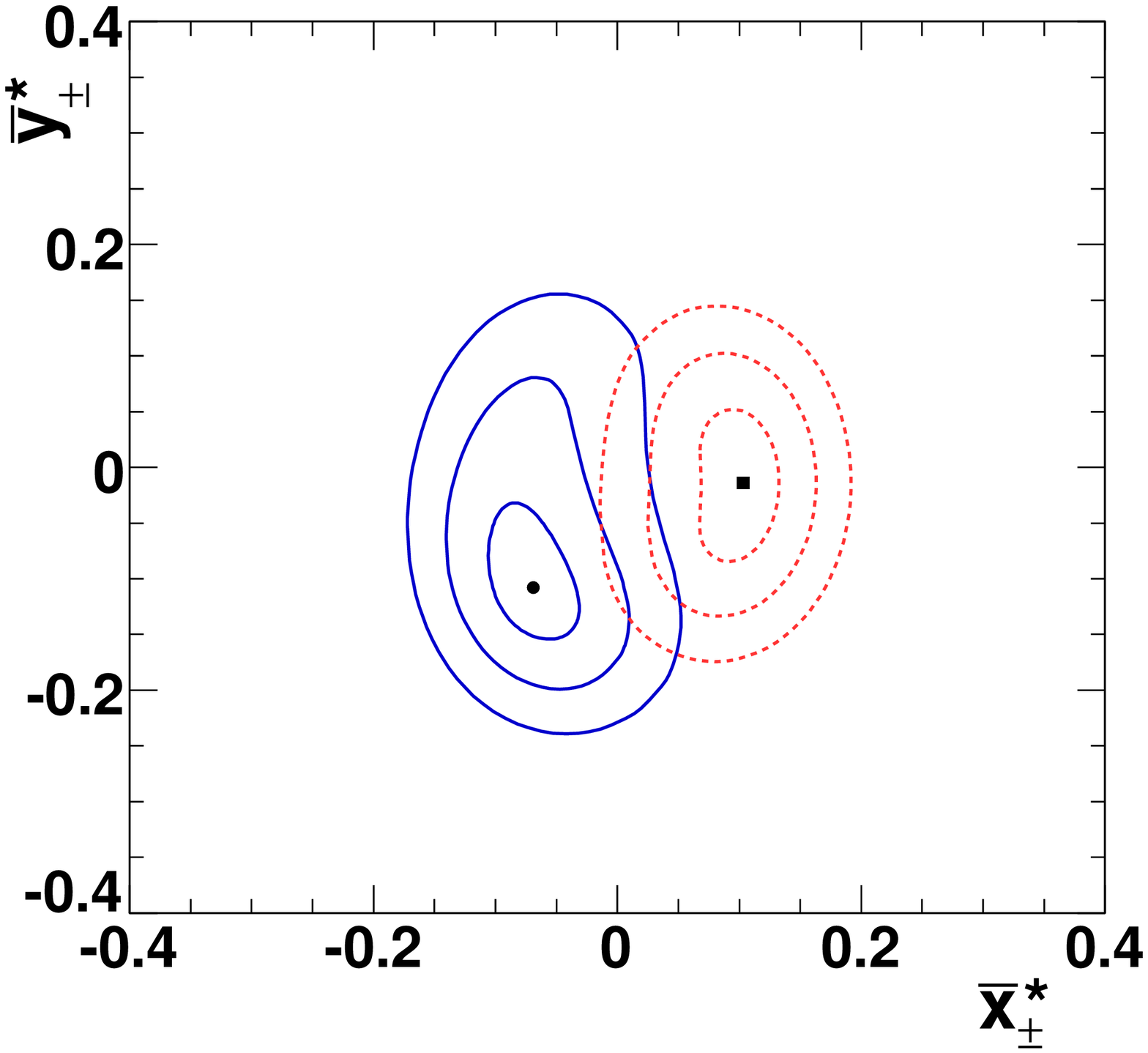,width=0.28\linewidth} &
  \epsfig{file=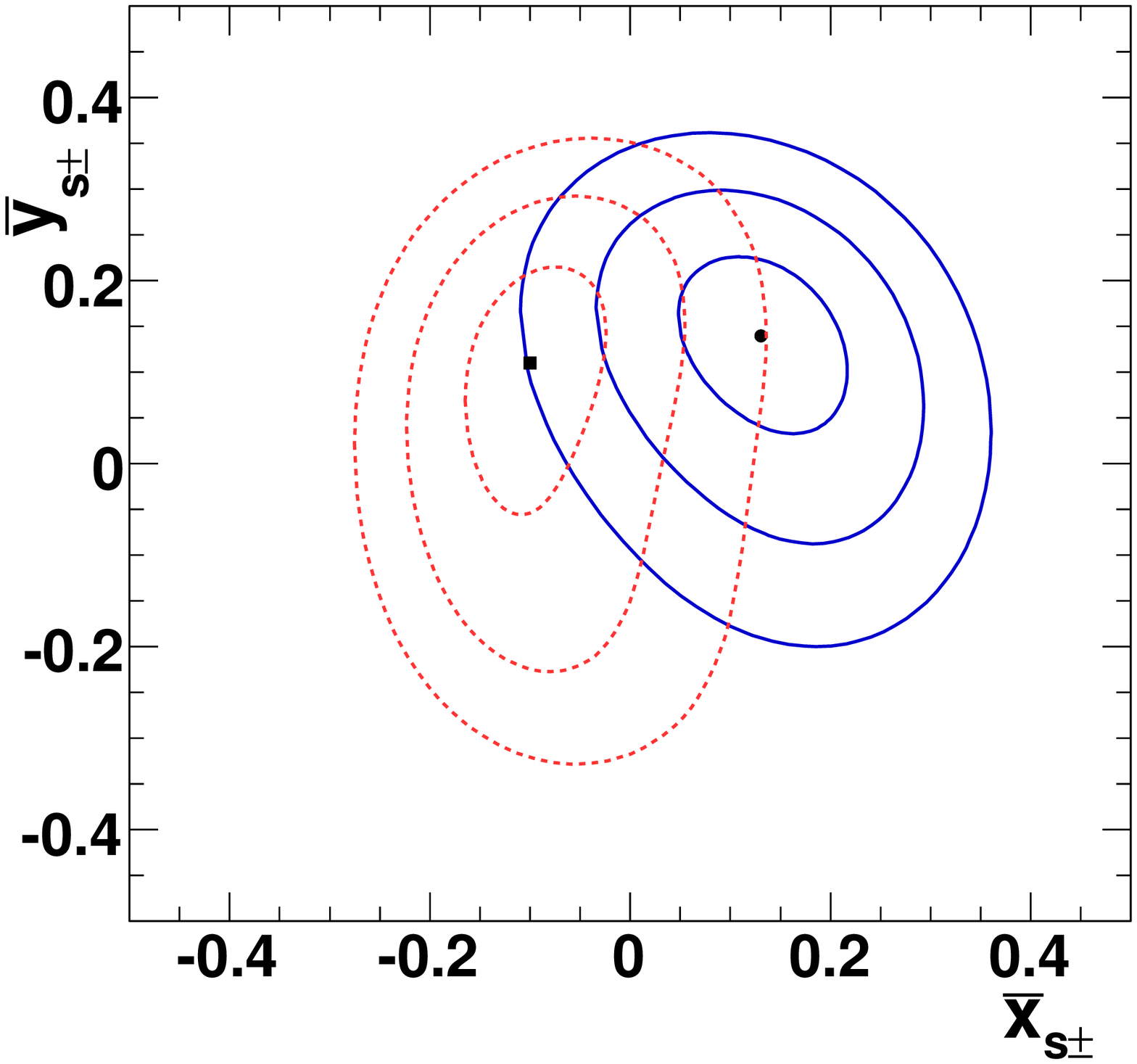,width=0.28\linewidth} \\
\end{tabular}
\caption{\label{fig:xyresults} 
(color online). Two-dimensional $-\twoDLL=s^2$ contours (up to three
standard deviations, i.e., $s=1,2,3$) in the 
\zbpmtrue (left column), \zbstpmtrue (center column), and \zspmtrue (right column) planes, for the GGSZ measurement only (top row),
the GGSZ and GLW combination (middle row), and the GGSZ, GLW, and ADS combination (bottom row).
The solid (blue) and dashed (red) lines correspond to \Bm and \Bp decays.
}
\end{center}
\end{figure*}

The results for the combined \CP-violating parameters \zbzbstpmtrue and \zspmtrue
are summarized in Table~\ref{tab:xyresults}. Figure~\ref{fig:xyresults} shows comparisons of
two-dimensional regions corresponding to \mbox{one-,} \mbox{two-,} 
and three-standard-deviation regions
in the \zbpmtrue, \zbstpmtrue, and \zspmtrue planes,
including statistical and systematic uncertainties,
for GGSZ only, GGSZ and GLW methods combined, and the overall combination.
These contours have been obtained using 
the likelihood ratio method, $-\twoDLL = s^2$, where $s$ is the number of standard deviations, where \twoDLL 
represents the variation of the combined log-likelihood with respect to its maximum value~\cite{ref:pdg2010}.
With this construction, the approximate confidence level (\CL) in two dimensions for each pair of variables is 
$39.3\%$, $86.5\%$, and $98.9\%$.
In these two-dimensional regions, the separation of the \Bm and \Bp positions is equal to $2\rb|\sin\g|$, $2\rbst|\sin\g|$, $2\krs|\sin\g|$ 
and is a measurement of direct \CP violation, while the angle between the lines connecting the \Bm and \Bp centers with the origin $(0,0)$ 
is equal to $2\g$.
Therefore, the net difference between \xbptrue and \xbmtrue observed in Table~\ref{tab:xyresults} and 
Fig.~\ref{fig:xyresults} is clear evidence for direct \CP violation in $\Bpm\to\D\Kpm$ decays.

In Fig.~\ref{fig:xyresults}, we observe that when the information from the GLW measurements is included
the constraints on the best fit values of the parameters are improved. 
However, the constraints on \ybpmtrue are poor due to the 
quadratic dependence and the fact that $\rb \ll 1$.
This is the reason why the GLW method alone can hardly constrain \g.
Similarly, Eq.~(\ref{eq:Rpm_ADS_vs_xy}) for the ADS method represents
two circles in the $(\xbpmtrue,\ybpmtrue)$ plane
centered at $(−\rb \cos\deltad, \rd\sin\deltad)$ and with radii $\sqrt{R_\pm}$.
It is not possible to determine \g with only ADS observables because the
true $(\xbpmtrue,\ybpmtrue)$ points are distributed over two circles~\cite{ref:rama_btodk}. 
Therefore, while the GLW and ADS methods
alone can hardly determine \g, when combined with the GGSZ measurements they help to improve significantly 
the constraints on the \CP-violating parameters \zbpmtrue, \zbstpmtrue, and \zspmtrue.

\begin{table}[htb] 
\caption{\label{tab:xyresults} 
\CP-violating complex parameters  
$\zbzbstpmtrue = \xbxbstpmtrue + i \ybybstpmtrue$ 
and  
$\zspmtrue = \xspmtrue + i \yspmtrue$ 
obtained from the combination of GGSZ, GLW, and ADS measurements.
The first error is statistical (corresponding to $-\twoDLL=1$), 
the second is the experimental systematic uncertainty including  
the systematic uncertainty associated to the GGSZ decay amplitude models.
} 
\begin{center} 
\begin{ruledtabular} 
\begin{tabular}{lrr} 
           & Real part (\%)\phzz\phzz &  Imaginary part (\%)\phzz \\ [0.025in] \hline  
 $\zbmtrue$    & $\phm8.1\pm2.3\pm0.7$    	     & $\phm4.4\pm3.4\pm0.5$ \\ 
 $\zbptrue$    & $-9.3\pm2.2\pm0.3$          	     & $-1.7\pm4.6\pm0.4$ \\ 
 $\zbstmtrue$  & $-7.0\pm3.6\pm1.1$                & $-10.6\pm5.4\pm2.0$    \\ 
 $\zbstptrue$  & $\phm10.3\pm2.9\pm0.8$             & $-1.4\pm8.3\pm2.5$ \\ 
 $\zsmtrue$    & $\phm13.3\pm8.1\pm2.6$    	     & $\phm13.9\pm8.8\pm3.6$    \\ 
 $\zsptrue$    & $-9.8\pm 6.9\pm1.2$               & $\phm11.0\pm 11.0\pm6.1$    \\ 
\end{tabular} 
\end{ruledtabular} 
\end{center} 
\end{table}

\section{ \bf Interpretation of Results}

In a second stage, we transform the combined $(\xbpmtrue,\ybpmtrue)$, $(\xbstpmtrue,\ybstpmtrue)$, and $(\xspmtrue,\yspmtrue)$
measurements into 
the physically relevant quantities \g and the set of hadronic parameters
$\uvec \equiv (\rb, \rbst, \krs, \deltab, \deltabst, \deltas)$.
We adopt a
frequentist procedure~\cite{ref:woodroofe} to obtain one-dimensional confidence intervals
of well-defined \CL that takes into account
non-Gaussian effects due to the nonlinearity of the relations between the observables and physical quantities.
This procedure is identical to that used in Refs.~\cite{ref:GGSZ2008,ref:GGSZ2010,ref:GLW_d0k,ref:GLWADS_d0kstar,ref:ADS_d0k_dstar0k_kpi}.

We define a $\chi^2$ function as 
\begin{eqnarray}
\chi^2(\g, \uvec) &\equiv& -\twoDLL(\g,\uvec)\nn\\
 &\equiv& -2 [ \ln{\cal L}(\g,\uvec) - \ln{\cal L}_{\mathrm{max}} ],~
\end{eqnarray}
where $\twoDLL(\g,\uvec)$ is the variation of the combined
log-likelihood with respect to its maximum value,
with the \zbzbstpmtrue and \zspmtrue expected values written in terms
of \g and \uvec, i.e., replacing \zbzbstpmtrue and \zspmtrue by $\rbrbst e^{i(\deltabdeltabst \pm \g)}$ 
and $\kappa \rs e^{i(\deltas \pm \g)}$, respectively.
To evaluate the \CL of a certain parameter (for example \g) at a given value ($\g_0$), we consider the value of 
the $\chi^2$ function at the new minimum, $\chi^2_\mathrm{min}(\g_0,\uvec_0)$, satisfying $\Delta \chi^2(\g_0) = \chi^2_\mathrm{min}(\g_0,\uvec_0) - \chi^2_\mathrm{min} \ge 0$.
In a purely Gaussian situation, 
the \CL is given by the probability that $\Delta \chi^2(\g_0)$ is exceeded for 
a $\chi^2$ distribution with one degree of freedom, $1-\CL = {\mathrm Prob}[\Delta \chi^2(\g_0);\nu=1]$, where $\mathrm{Prob}[\Delta \chi^2(\g_0);\nu=1]$ is 
the corresponding cumulative distribution function (this approach is later referred to as ``Prob method'')~\cite{ref:pdg2010}. 
In a non-Gaussian situation one has to consider $\Delta \chi^2(\gamma_0)$ as a test statistic, and 
rely on a Monte Carlo simulation to obtain its expected distribution. 
This Monte Carlo simulation is performed by
generating 
more than $10^9$ samples 
(sets of the 39 GGSZ, GLW, ADS, and \D decay observable values), using the combined likelihood 
evaluated at values $(\g_0,\uvec_0)$, i.e., ${\cal L}(\g_0,\uvec_0)$.
The confidence level $\CL$ is determined from the fraction of experiments for which
$\Delta \chi'^2(\g_0) > \Delta \chi^2(\g_0)$,
where $\Delta \chi'^2(\g_0) = \chi'^2(\g_0,\uvec'_0) - \chi'^2_{\mathrm{min}}$ for each simulated experiment 
is determined as in the case of the actual data sample. 
We adopt the Monte Carlo simulation method as baseline to determine the
\CL, and allow $0 \le \rbrbst,\krs \le 1$ and $-180^\circ \le \g, \deltabdeltabst,\deltas \le 180^\circ$.

\begin{figure}[htb!]
\begin{center}
\epsfig{file=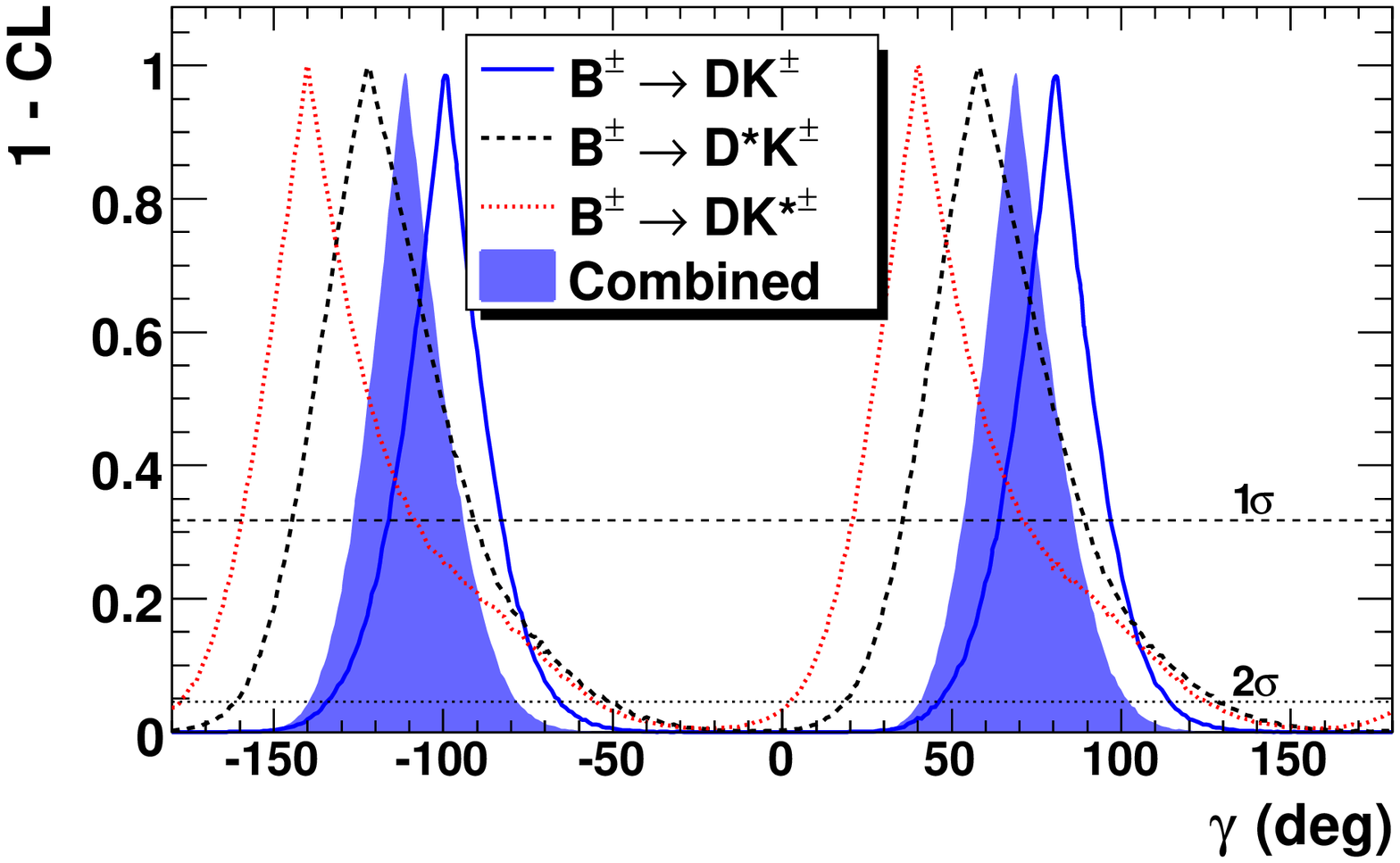,width=0.85\linewidth} \\
\epsfig{file=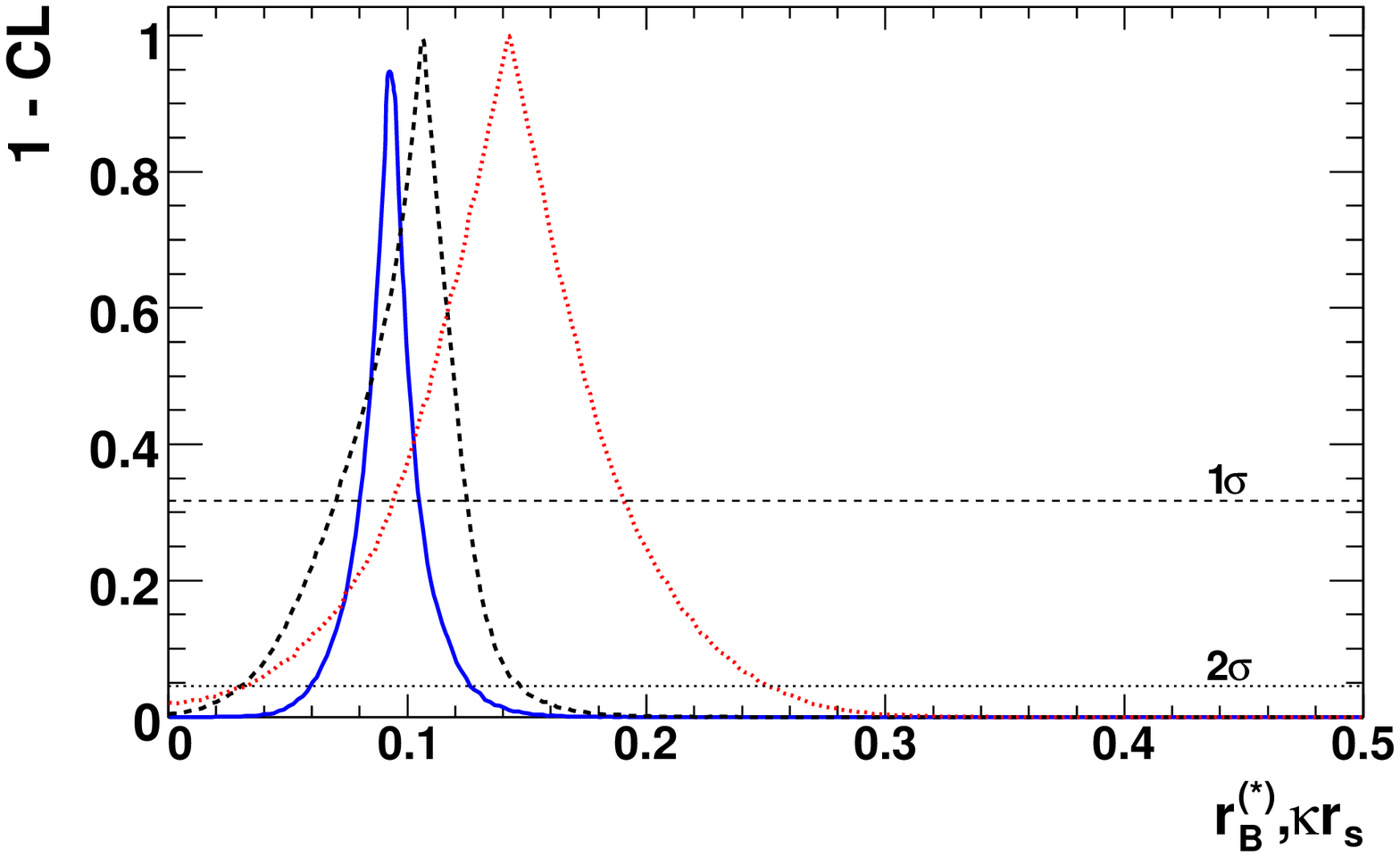,width=0.85\linewidth} \\
\epsfig{file=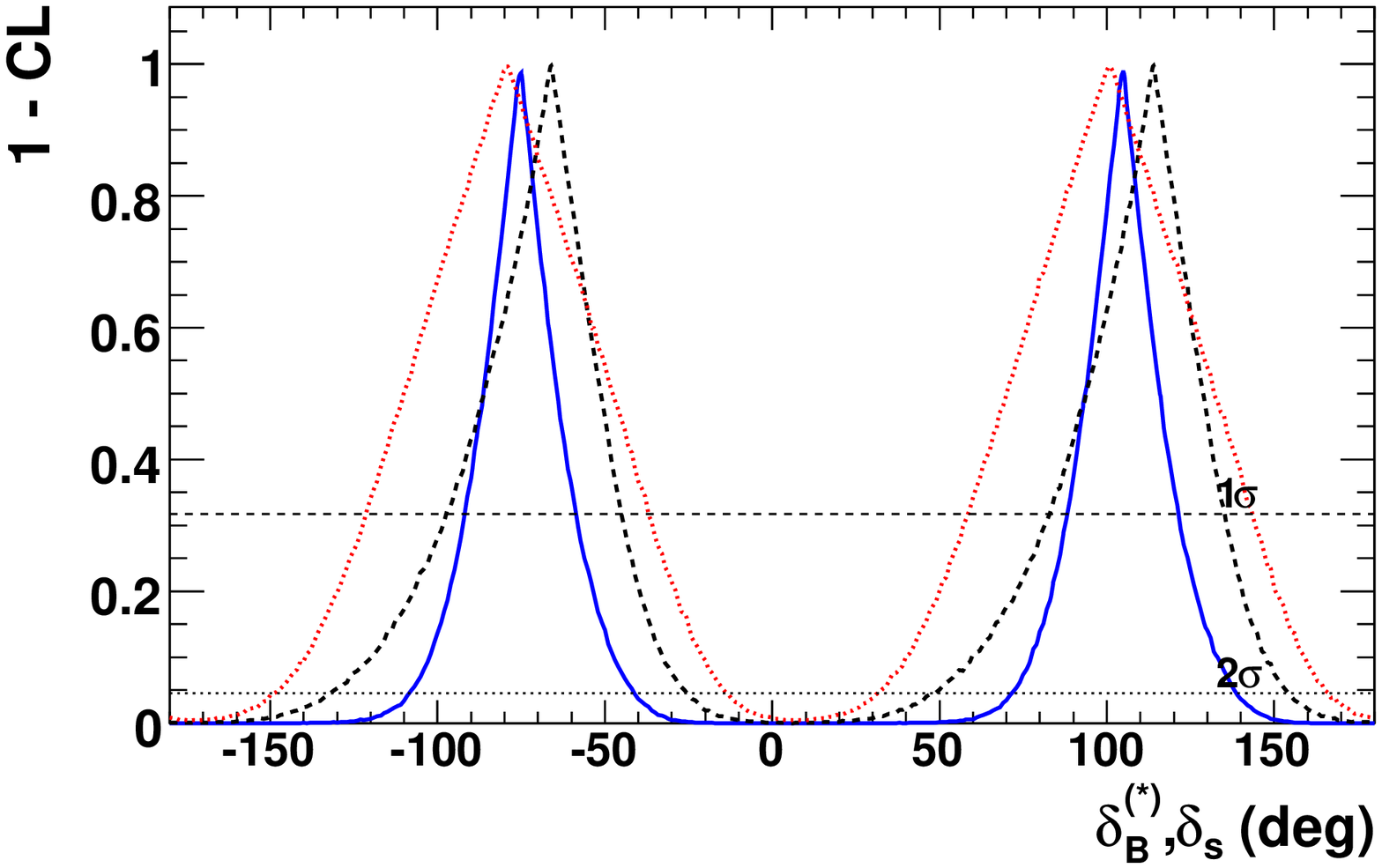,width=0.85\linewidth}
\end{center}
\caption{(color online). 
$1-\CL$ distributions for the combination of the GGSZ, GLW and ADS methods
as a function of \g (top), \rbrbst, and \krs (middle), and \deltabdeltabst, \deltas (bottom),
including statistical and systematic uncertainties,
for $\Bpm \to \D \Kpm$, $\Bpm \to \Dstar \Kpm$, and $\Bpm \to \D \Kstarpm$ decays. The combination of all the \B decay channels is also shown for \g.
The dashed (dotted) horizontal line corresponds to the one- (two-) standard-deviation \CL.
\label{fig:scans-gamma-rb-delta}
}
\end{figure}

\begin{figure}[htb!]
\begin{center}
\begin{tabular} {c}
\epsfig{file=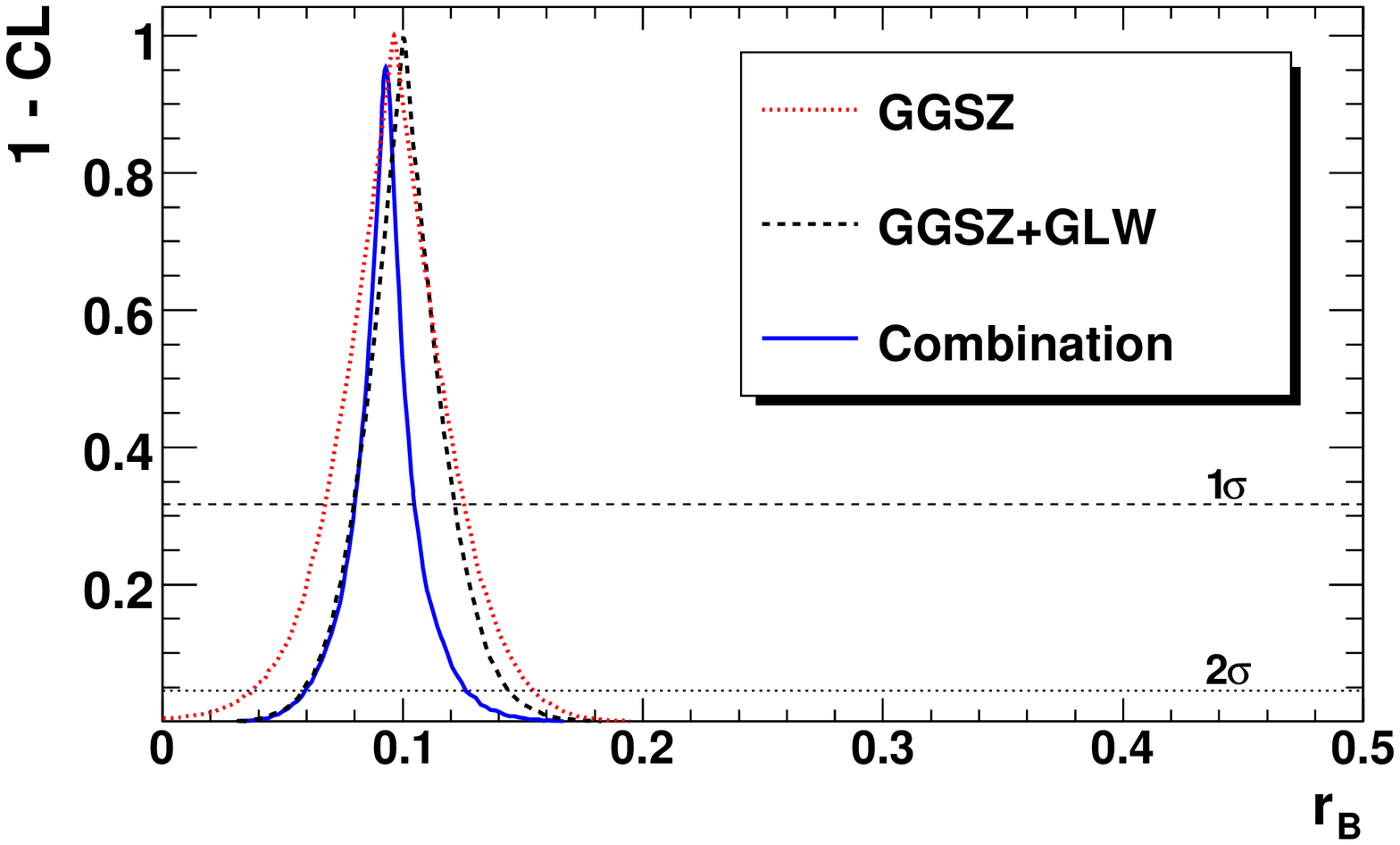,width=0.85\linewidth} \\
\epsfig{file=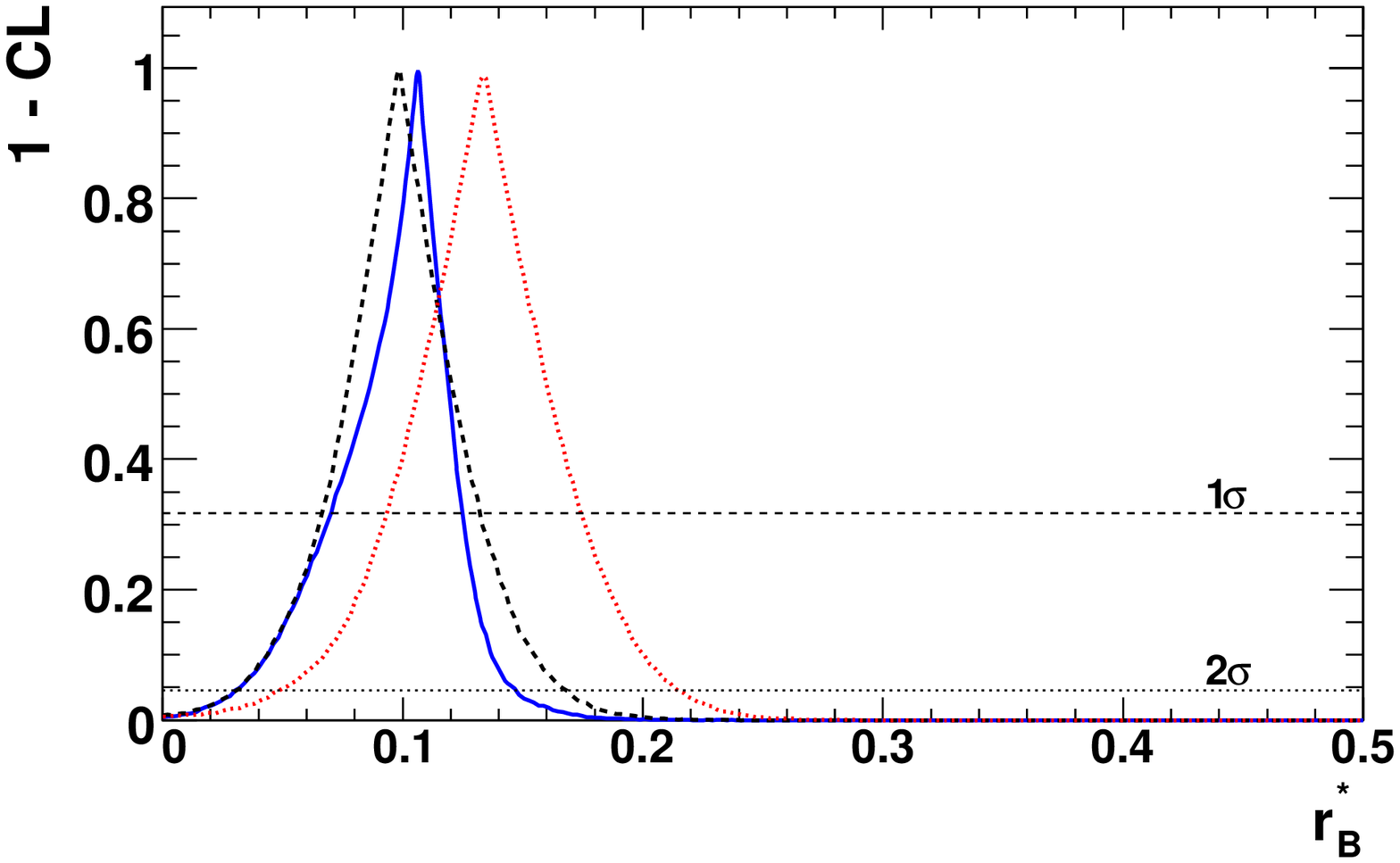,width=0.85\linewidth} \\
\epsfig{file=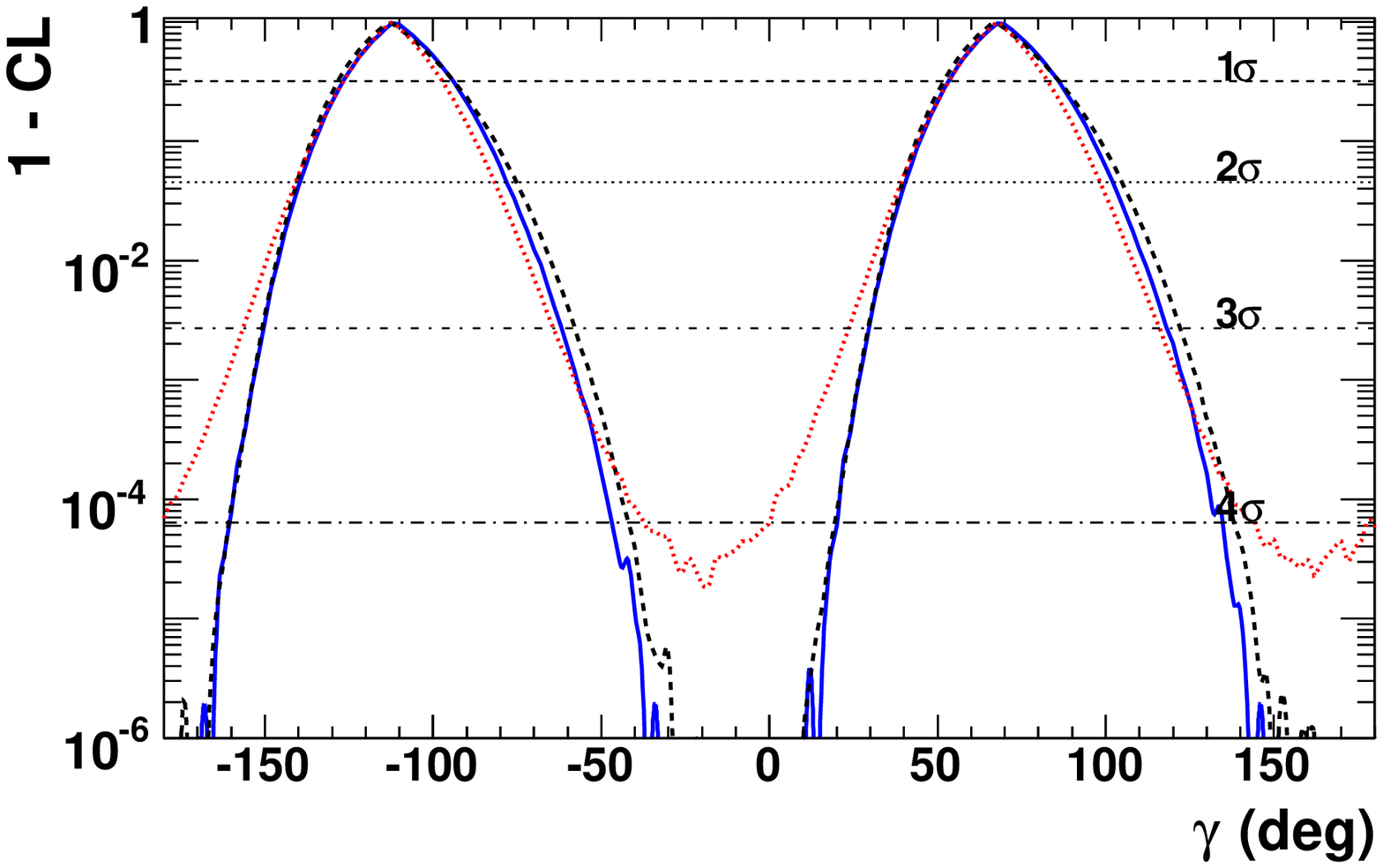,width=0.85\linewidth} 
\end{tabular}
\end{center}
\caption{(color online). Comparison of $1-\CL$ as a function of \rb (top), \rbst (middle), and \g (bottom) for all \B decay channels combined with the 
GGSZ method only, the combination with the GLW measurements, and the global combination, 
including statistical and systematic uncertainties.
The horizontal lines represent the one-, two-, three- and four-standard-deviation \CL.
\label{fig:scan-comp-rb+gamma}
}
\end{figure}

Figure~\ref{fig:scans-gamma-rb-delta} illustrates $1-\CL$ as a function of \g, \rbrbst, \krs, \deltabdeltabst, and \deltas,
for each of the three \B decay channels separately and, in the case of \g, their combination.
The combination has the same twofold ambiguity in the weak and strong phases
as that of the GGSZ method,
$(\gamma;\deltabdeltabst,\deltas) \to (\gamma + 180^\circ; \deltabdeltabst + 180^\circ,\deltas+180^\circ)$.
From these distributions, we extract one- and two-standard-deviation intervals as the sets of values for which
$1-\CL$ is greater than 31.73\% and 4.55\%, respectively, as summarized in Table~\ref{tab:polarresults}.
When comparing these intervals to those obtained with the GGSZ method only, also shown in Table~\ref{tab:polarresults},
we observe that the combination helps improving the constraints on \rbrbst and \krs, but not those on \g.
To assess the impact of the GLW and ADS observables in the determination of \g, we compare $1-\CL$ as a function of \rbrbst and \g 
for all \B decay channels combined using the GGSZ method alone, the combination with the GLW measurements, and the global combination, 
as shown in Fig.~\ref{fig:scan-comp-rb+gamma}. While the constraints on \rb are clearly 
improved at the one- and two-standard-deviation level, and to a lesser extent on \rbst, their best (central) values move towards slightly
lower values. Since the uncertainty on \g scales approximately as $1/\rbrbst$,
the constraints on \g at $68.3\%$ and $95.4\%$ \CL do not improve,
in spite of the tighter constraints on the combined measurements shown in Fig.~\ref{fig:xyresults}.
However, adding GLW and ADS information reduces the confidence intervals for smaller $1-\CL$, as a consequence of
the more Gaussian behavior when the significance of excluding $\rbrbst=0$ increases.
Thus, for example, in the region close to four standard deviations, the GGSZ method alone 
does not constrain \g, while the combination is able to exclude large regions.


\begin{table}[htb]
\caption{\label{tab:polarresults} $68.3\%$ and $95.5\%$ 1-dimensional \CL regions,
equivalent to one- and two-standard-deviation intervals,
for \g, \deltabdeltabst, \deltas, \rbrbst, and \krs,
including all sources of uncertainty, obtained from the combination of GGSZ, GLW, and ADS measurements.
The combined results are compared to those obtained using the GGSZ measurements only, taken from Ref.~\cite{ref:GGSZ2010}.
The results for \g, \deltabdeltabst, and \deltas\ are given modulo 
a $180^\circ$ phase.
}
\begin{center}
\begin{ruledtabular}
\begin{tabular}{lrrrr}
Parameter            & \multicolumn{2}{c}{$68.3\%$ \CL}  & \multicolumn{2}{c}{$95.5\%$ \CL} \\ 
                     & Combination & GGSZ & Combination & GGSZ \\ [0.035in] \hline
 \noalign{\vskip3pt}
\g $(^\circ)$         & $69^{+17}_{-16}$       &  $68^{+15}_{-14}$     & $[41,102]$   & $[39,98]$ \\
\rb $(\%)$           & $9.2 ^{+1.3}_{-1.2}$   &  $9.6 \pm 2.9$      & $[6.0,12.6]$ & $[3.7,15.5]$ \\
\rbst $(\%)$         & $10.6^{+1.9}_{-3.6}$   &  $13.3^{+4.2}_{-3.9}$  & $[3.0,14.7]$ & $[4.9,21.5]$ \\
\krs $(\%)$          & $14.3^{+4.8}_{-4.9}$   &  $14.9^{+6.6}_{-6.2}$  & $[3.3,25.1]$ & $< 28.0$ \\
\deltab   $(^\circ)$  & $105^{+16}_{-17}$      &  $119^{+19}_{-20}$    & $[72,139]$   & $[75,157]$ \\
\deltabst $(^\circ)$  & $-66^{+21}_{-31}$      &  $-82\pm21$        & $[-132,-26]$ & $[-124,-38]$ \\
\deltas   $(^\circ)$  & $101\pm43$          &  $111\pm32$        & $[32,166]$   & $[42,178]$ \\
\end{tabular}
\end{ruledtabular}
\end{center}
\end{table}

The significance of direct \CP violation is obtained by evaluating $1-\CL$ for the most probable \CP conserving 
point, i.e., the set of hadronic parameters \uvec with $\g=0$. Including 
statistical and systematic uncertainties, we obtain $1-\CL=3.4\times 10^{-7},\ 2.5\times 10^{-3}$, and $3.6\times 10^{-2}$,
corresponding to $5.1$, $3.0$, and $2.1$ standard deviations, for
$\Bpm\to\D\Kpm$, $\Bpm\to\Dstar\Kpm$, and $\Bpm\to\D\Kstarpm$ decays, respectively.
For the combination of the three decay modes we obtain $1-\CL=3.1\times 10^{-9}$, corresponding to $5.9$ standard deviations.
For comparison,
the corresponding significances with the GGSZ method alone are
$2.9$, $2.8$, $1.5$, and $4.0$ standard deviations~\cite{ref:CPsignificanceFix},
while with the GGSZ and GLW combination they are
$4.8$, $2.7$, $1.8$, and $5.4$, respectively.

The frequentist procedure used to obtain \g and the hadronic parameters \uvec
is not guaranteed to have perfect coverage, especially 
for low values of \rbrbst, \rs.
This is due to the treatment of nuisance
parameters~\cite{ref:woodroofe}.
Instead of scanning the entire parameter
space defined by \g and \uvec (seven dimensions), we perform one-dimensional scans,
in which, during MC generation, the nuisance parameters are set to their re-optimized
best-fit values at each scan point. 
In order to evaluate the coverage properties of our procedure, we generate 
more than $10^9$ samples 
with true values of $(\g,\uvec)$ set to their
best fit values, $(\g_{\mathrm{best}},\uvec_{\mathrm{best}})$,
as given in Table~\ref{tab:polarresults}. For each generated experiment, we determine
$1-\CL'$ at $\g_0 = \g_{\mathrm{best}}$, as done previously with the actual data sample 
using the Monte Carlo simulation method. 
The statistical coverage $\alpha$, defined as the probability for the true value of \g ($\g_0$) to be inside the given $1-\CL$ interval, 
is evaluated as the fraction of experiments with $1-\CL'$ larger than $1-\CL$. We obtain $\alpha=0.679\pm0.005\ (0.955\pm0.002)$
for the combination, and $\alpha = 0.670\pm0.005\ (0.950\pm0.002)$ for the GGSZ method alone, for $\CL=0.683\ (0.954)$,  respectively.
For comparison purposes, the corresponding values using the Prob method are $\alpha=0.641\pm0.005\ (0.941\pm0.003)$ and $\alpha = 0.609\pm0.005\ (0.920\pm0.003)$.
While the Prob method tends to underestimate the confidence intervals, the Monte Carlo simulation method provides intervals with correct coverage, especially
for the combination where the magnitude ratios between the suppressed and favored decays have more stringent constraints.

\section{ \bf Summary}

In summary, using up to $474\times 10^6$ \BB decays recorded by the \babar
detector, we have presented a combined measurement of the
\CP-violating 
ratios between the $\b\to \u \cbar \s$ and $\b \to \c \ubar \s$ amplitudes
in processes $\Bpm\to\DDstar\Kpm$ and
$\Bpm\to\D\Kstarpm$. The combination procedure maximizes the information provided by the most sensitive \g measurements and analysis techniques that exploit 
a large number of \D decay final states, including three-body
self-conjugate, \CP, and doubly-Cabibbo-suppressed states, 
resulting in the most precise measurement of these ratios.
From the measurements of these ratios we determine $\g =
(69^{+17}_{-16})^\circ$ (modulo $180^\circ$), where the total 
uncertainty is dominated by the statistical component, with 
the experimental and amplitude model systematic uncertainties
amounting to $\pm 4^\circ$.
We also derive the most precise determinations of the magnitude
ratios \rbrbst and \krs. 
The two-standard-deviation region for $\gamma$ is $41^\circ < \gamma
< 102^\circ$. 
The combined significance of $\g \ne 0$ is $1-\CL=3.1\times 10^{-9}$,
corresponding to $5.9$ standard deviations, meaning
observation of direct \CP violation in the measurement of \g. 
These results supersede our previous 
constraints based on the GGSZ, GLW, and ADS analyses of 
charged \B decays~\cite{ref:GGSZ2010,ref:GGSZ2008,ref:GGSZ2005,ref:GLW_d0k,ref:GLW_dstar0k,ref:ADS_d0k_dstar0k_kpi,ref:ADS_d0k_kpipi0},
and are consistent with the range of values implied by other 
experiments~\cite{ref:belle_GGSZ_ModelIndep,ref:belle_GGSZ,ref:lhcb_GGSZ,ref:belle_glw,ref:cdf_glw,ref:lhcb_glw,ref:belle_ads,ref:belle_ads_DKstar,ref:cdf_ads,ref:lhcb_glw}.

\par
\begin{center}
{\small \bf ACKNOWLEDGMENTS}
\end{center}
We are grateful for the 
extraordinary contributions of our \pep2\ colleagues in
achieving the excellent luminosity and machine conditions
that have made this work possible.
The success of this project also relies critically on the 
expertise and dedication of the computing organizations that 
support \babar.
The collaborating institutions wish to thank 
SLAC for its support and the kind hospitality extended to them. 
This work is supported by the
US Department of Energy
and National Science Foundation, the
Natural Sciences and Engineering Research Council (Canada),
the Commissariat \`a l'Energie Atomique and
Institut National de Physique Nucl\'eaire et de Physique des Particules
(France), the
Bundesministerium f\"ur Bildung und Forschung and
Deutsche Forschungsgemeinschaft
(Germany), the
Istituto Nazionale di Fisica Nucleare (Italy),
the Foundation for Fundamental Research on Matter (The Netherlands),
the Research Council of Norway, the
Ministry of Education and Science of the Russian Federation, 
Ministerio de Econom\'{\i}a y Competitividad (Spain), and the
Science and Technology Facilities Council (United Kingdom).
Individuals have received support from 
the Marie-Curie IEF program (European Union) and the A. P. Sloan Foundation (USA).


\begin{thebibliography}{99}
\bibitem{ref:charge} Charge conjugate modes are implicitly included unless otherwise stated.
\end{thebibliography}
\end{document}